\documentclass[aps,prd,a4paper,longbibliography,reprint,showpacs,superscriptaddress,nofootinbib,amsmath,amssymb]{revtex4-1}
\usepackage{graphicx}
\usepackage[latin2]{inputenc}
\usepackage{dsfont}
\usepackage{mathrsfs}
\usepackage{slashed}
\usepackage{array}
\usepackage{colordvi}
\usepackage{rotfloat}
\usepackage{rotating}
\usepackage{color}
\usepackage[    bookmarks,
                 bookmarksopen = true,
                 bookmarksnumbered = true,
                 linktocpage,
                 colorlinks = true,
                 linkcolor = blue,
                 urlcolor  = blue,
                 citecolor = blue,
                 anchorcolor = green,
                 hyperindex = true,
                 hyperfigures]
		{hyperref}
\usepackage{cmap}
\usepackage{xcolor}

\newcommand{\Laplace}{\mathop{{}\bigtriangleup}\nolimits}

\newcommand{\be}{\begin{equation}}
\newcommand{\ee}{\end{equation}}

\newcommand{\tr}{\mathop{\textmd{tr}}\nolimits}
\newcommand{\re}{\mathop{\textmd{Re}}\nolimits}

\newcommand{\units}[1]{\ensuremath{\,\textmd{#1}}}
\newcommand{\bra}[1]{\left\langle #1 \right|}
\newcommand{\ket}[1]{\left| #1 \right\rangle}

\newcommand{\vect}[1]{\ensuremath{\boldsymbol{#1}}}
\newcommand{\wupc}{\ensuremath{\varepsilon}}
\newcommand{\apec}{\ensuremath{\delta}}
\newcommand{\unitj}{\ensuremath{\hat{\boldsymbol{\jmath}}}}
\newcommand{\uniti}{\ensuremath{\hat{\boldsymbol{\imath}}}}

\newcommand{\oper}{\ensuremath{\mathcal{O}}}
\newcommand{\GC}{\ensuremath{G}}

\begin{document}
\title{Novel quark smearing for hadrons with high momenta in lattice QCD}
\author{Gunnar S.~Bali}
  \email{gunnar.bali@ur.de}
\affiliation{Institut f\"ur Theoretische Physik, Universit\"at Regensburg, 93040 Regensburg, Germany}
\affiliation{Department of Theoretical Physics, Tata Institute of Fundamental Research, Homi Bhabha Road, Mumbai 400005, India}
\author{Bernhard Lang}
  \email{bernhard.lang@physik.uni-regensburg.de}
\affiliation{Institut f\"ur Theoretische Physik, Universit\"at Regensburg, 93040 Regensburg, Germany}
\author{Bernhard U.~Musch}
  \email{bmusch@jlab.org}
\affiliation{Institut f\"ur Theoretische Physik, Universit\"at Regensburg, 93040 Regensburg, Germany}
\author{Andreas Sch\"afer}
\affiliation{Institut f\"ur Theoretische Physik, Universit\"at Regensburg, 93040 Regensburg, Germany}
\collaboration{RQCD Collaboration}
\date{\today}
\begin{abstract} 
Hadrons in lattice QCD are usually created employing smeared interpolators.
We introduce a new quark smearing that
allows us to maintain small statistical errors and good overlaps of hadronic
wave functions with the respective ground states, also at high spatial
momenta. The method is successfully tested for the pion and the nucleon
at a pion mass $m_{\pi}\approx 295\units{MeV}$ and momenta as high as
2.8\units{GeV}. We compare the results obtained to dispersion
relations and suggest further optimizations.
\end{abstract}
\maketitle
\section{Introduction}
Lattice QCD simulations predict an ever increasing number of observables
that are relevant to particle and hadron physics phenomenology.
These results are usually extracted from expectation values
of $n$-point functions at large Euclidean time separations.
Due to the decay of these functions with time, statistical noise
over signal ratios increase exponentially as time separations
are taken large (with the notable exception of pseudoscalar
mesons at zero momentum).
Fortunately, there exists some freedom in the construction of interpolators
for the creation of mesonic and baryonic states. Employing interpolators
that resemble the spatial structure of
the ground state wave function enables asymptotic results to be extracted at
time separations where the signal over noise ratio is still large.

Many applications nowadays demand hadrons that carry momentum.
For instance, pushing the calculation of semileptonic decay form
factors for $B\rightarrow\pi\ell\bar{\nu}_{\ell}$ or
$\Lambda_b\rightarrow p\ell\bar{\nu}_{\ell}$~\cite{Detmold:2015aaa}
towards small virtualities requires spatial momenta of the
size of the mass difference between the $B$ meson and
the pion or between the $\Lambda_b$ baryon and the proton, respectively.
Another important type of application are parton distribution functions (PDFs) and their generalizations, in particular transverse momentum dependent parton distribution functions (TMDs), or Wigner distributions as a whole. 
Also these quantities are extracted from matrix elements of the type $ \bra{H(\vect{p}')} \oper \ket{H(\vect{p})}$, where $H(\vect{p})$
is a hadron state with momentum $\vect{p}$.
The operator $\oper$ cannot have an extent in Minkowski-time on the Euclidean
lattice. Therefore, in order to extrapolate to light front kinematics for an inherently non-local operator,
one has to work in a frame of reference where hadrons carry high spatial momenta $\vect{p}$, $\vect{p}'$.
This fact has been known for quite some time in the context of lattice calculations of TMDs \cite{Hagler:2009mb,Musch:2010ka,Musch:2011er} 
and is illustrated very clearly in recent work on these distributions in the pion~\cite{Engelhardt:2015xja}.
For the same reason fast hadrons on the lattice are highly desirable in a new scheme proposed 
to relate quasi parton distributions to light front
distributions in a controlled manner \cite{Ji:2013dva,Ji:2015jwa}.
First lattice
computations in this direction have
started~\cite{Lin:2014zya,Alexandrou:2015rja}.
Earlier suggestions to compute quasi distribution amplitudes
in position space~\cite{Aglietti:1998ur,Abada:2001if,Braun:2007wv}
equally require
pions or nucleons at high momenta. Unfortunately,
up to now no satisfactory techniques for hadrons carrying high momenta
existed to suppress excited
state contributions.

To be more specific, a two-point function
is given as
\begin{equation}
\label{eq:twopoint}
C_H(t)=\langle H(t)H^{\dagger}(0)\rangle
=\sum_{j>0}\frac{|\langle j|\hat{H}^{\dagger}|0\rangle|^2}{2E_{H,j}}e^{-E_{H,j}t}\,,
\end{equation}
where $E_{H,j}$ denotes the $j$th energy level within the tower
of states created by the interpolator
$H^{\dagger}$. Obviously, the contribution of the $j$th excited
state is suppressed relative to the ground state
not only by $\exp[-(E_{H,j}-E_{H,1})t]$ but also
by the ratio
$|\langle j|\hat{H}^{\dagger}|0\rangle|^2/|\langle 1|\hat{H}^{\dagger}|0\rangle|^2$:
increasing the ground state overlap factor
$|\langle 1|\hat{H}^{\dagger}|0\rangle|$,
relative to $|\langle j|\hat{H}^{\dagger}|0\rangle|$ for $j>1$,
results in an additional suppression of excitations.

Reducing excited state overlaps by employing extended
interpolators was first pursued in computations of the
glueball spectrum. In this case the gauge links within the corresponding
interpolators can be iteratively ``APE smeared''~\cite{Falcioni:1984ei},
``fuzzed''~\cite{Teper:1987wt} or ``HYP smeared''~\cite{Hasenfratz:2001hp},
to better approximate the (smooth) ground state wave function, see
also Ref.~\cite{Gupta:1990ek}.
This gauge link smearing was subsequently generalized
to iterative smearing of quark fields within interpolators
that create mesonic and baryonic states, in particular gauge covariant Wuppertal
(i.e.~Gau\ss{}) smearing~\cite{Gusken:1989ad,Gusken:1989qx,Alexandrou:1990dq},
hydrogen-like smearing~\cite{Alexandrou:1990dq}
as well as Jacobi smearing~\cite{Collins:1992rp,Allton:1993wc}.
Additionally, in Refs.~\cite{Gusken:1989qx,Alexandrou:1990dq,Bali:2005fu}
APE smearing was employed for the spatial gauge transporters within
the quark smearing while in Ref.~\cite{Bali:2005fu}
linear combinations of different levels of Wuppertal smearing
were utilized.

Since Gaussian smearing functions may not be optimal for creating, e.g.,
$p$-waves, even when adding derivatives to the interpolator,
iterative smearing was later-on combined with
displaced quark sources (fuzzing) in Ref.~\cite{Lacock:1994qx},
a generalization of which was suggested in Ref.~\cite{Boyle:1999gx}.
Finally, in Ref.~\cite{vonHippel:2013yfa} ``free form smearing'',
folding Gaussian smearing with an arbitrary function in a gauge
covariant way, was invented. Preceding and in parallel to
gauge covariant iterative smearing functions, gauge fixed sources have
been utilized: wall sources for zero~\cite{Bacilieri:1988fh} and
non-zero momentum~\cite{Detmold:2012wc}, box~\cite{Allton:1990qg}
sources, Gaussian ``shell sources''~\cite{DeGrand:1990dz}
and sources with nodes~\cite{DeGrand:1991ng}.
These gauge fixed methods and free form smearing share the disadvantage
that smearing the sink requires all quark positions to
be summed over individually, turning this prohibitively expensive.
Having identical source and sink interpolators,
however, is very desirable as only this guarantees the
positivity of the coefficients of the spectral decomposition
Eq.~\eqref{eq:twopoint} and thus the convexity of two-point functions.
For completeness, we also mention the ``distillation'' (or
Laplacian-Heaviside) method of Ref.~\cite{Peardon:2009gh}
since this is closely related to gauge covariant smearing.

Large momenta increase the energy of the state and result in faster
decaying two- and three-point functions and, therefore, in inferior noise to
signal ratios. Moreover, as we shall see,
excited state suppression becomes far less effective when
using conventional quark smearing methods.
Some attempts have been
made~\cite{Roberts:2012tp,DellaMorte:2012xc}
to introduce an anisotropy into Wuppertal
smearing~\cite{Gusken:1989ad,Gusken:1989qx}, aiming at
Lorentz contracting the interpolating wave function
according to the boost factor $1/\gamma=m/E(\vect{p})$,
along the direction of the spatial
momentum $\vect{p}$. However,
this did not result in the ground state enhancement that one would have
hoped for. Here we will argue and demonstrate that to achieve satisfactory
results at high momenta, additional phase factors need to be incorporated
into quark smearing functions.

This article is organized as follows.
First, in Sec.~\ref{sec:basic}, we discuss the basic idea behind the new
class of smearing functions that we introduce.
Then, in Sec.~\ref{sec:wuppertal} we are more specific, modifying
Wuppertal smearing as a generic example and suggest further improvements.
In Sec.~\ref{sec:simulate} we discuss our simulation
parameters and expectations for
the nucleon and pion energies. After the stage is set,
in Sec.~\ref{sec:results} we investigate the feasibility of the method
in a realistic numerical study, optimize the smearing parameters and pursue a
comparison between the new and the conventional method. Finally,
we study the pion and nucleon dispersion relations, before we conclude.

\section{Momentum smearing: the basic idea}
\label{sec:basic}
\begin{figure}
\includegraphics[width=0.48\textwidth]{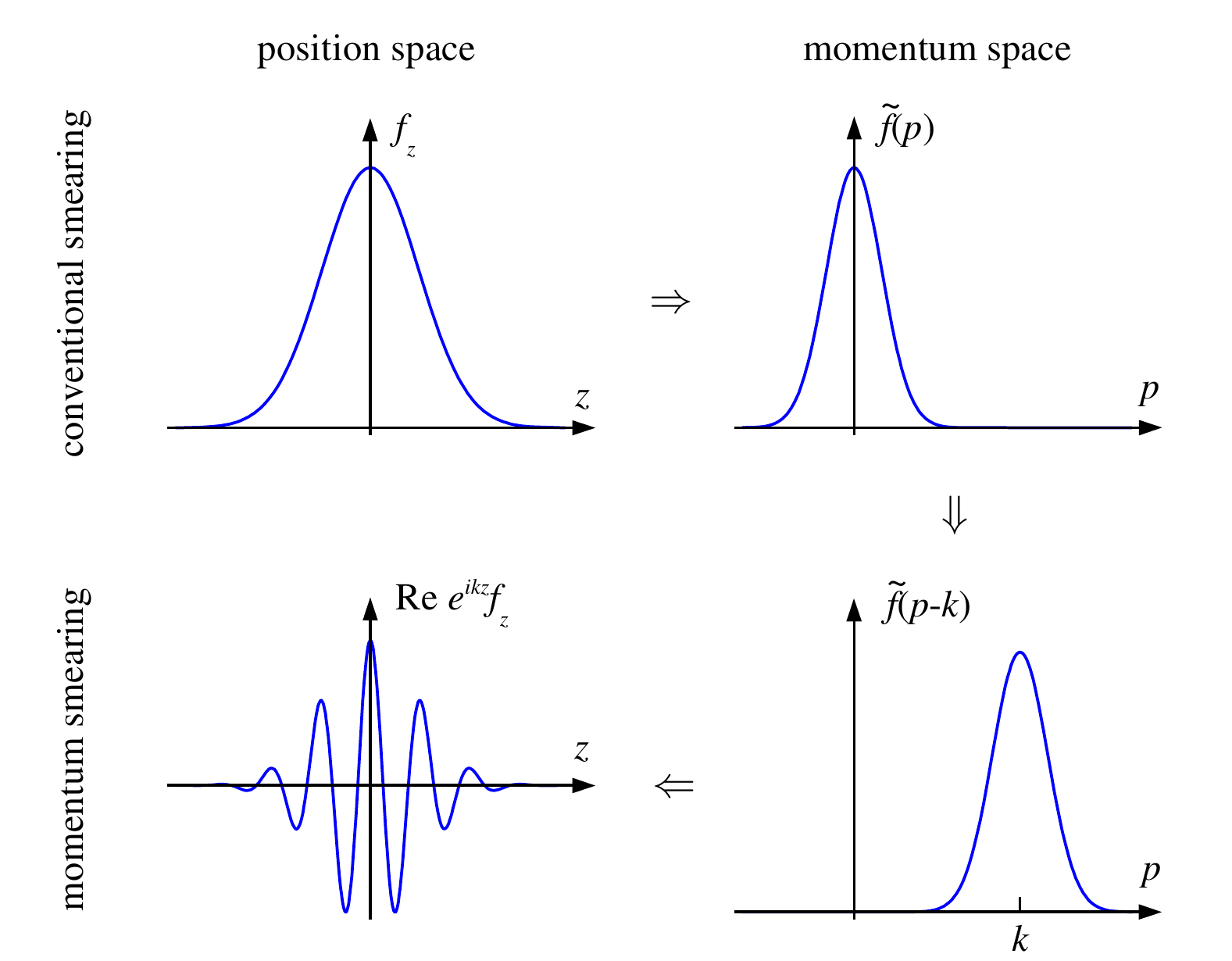}
\caption{Conventional smearing versus momentum smearing for the example
of a Gaussian wave function in $d=1$ spatial dimensions.
The momentum $k$ shifts the centre of the distribution
in momentum space, resulting in an oscillatory behaviour in position space.}
\label{fig:transform}
\end{figure}

As discussed above, quark smearing within hadronic sources or sinks is essential in lattice simulations
to increase the overlap with the desired physical state, reflecting the fact that hadrons are extended objects,
rather than pointlike. A smearing operator $F$ is diagonal in time, trivial
in spin and acts
on the position and colour indices of quark fields:
\begin{equation}
	(F q)_{\vect{x}} = \sum_{\vect{y}\in (a \mathbb{Z})^d}\, f_{\vect{x}-\vect{y}}\, \GC_{\vect{x}\vect{y}}\, q_{\vect{y}} \label{formwupN}\,,
\end{equation}
where $f$ is a scalar function, $\GC$ is a gauge covariant transporter,
which in the free case will be a unit matrix in colour and position space,
and $d$ is the number of spatial dimensions, usually $d=3$.
Note that the field $q_{\vect{x}}$ is usually periodic in $\vect{x}$ on the lattice, whereas $f_{\vect{x}-\vect{y}}$ need not be periodic in $\vect{x}-\vect{y}$. 
In the free case, the convolution Eq.~\eqref{formwupN} becomes a product in Fourier space
\begin{equation}
	\sum_{\vect{x}\in (a \mathbb{Z})^d} e^{i\vect{p}\cdot\vect{x}}\,(F q)_{\vect{x}} = 
	\tilde{f}(\vect{p})\, \tilde{q}_{\vect{p}}\,.	\label{wupfour}
\end{equation}
For the special case of a Gaussian,
\begin{equation}
\label{eq:smgauss}
	 f_{\vect{x}-\vect{y}} = f_{\vect{0}} \exp\left(-\frac{|\vect{x}-\vect{y}|^2}{2{\sigma}^2}\right)\,,
\end{equation}
the Fourier transformed smearing kernel again is a Gaussian:
\begin{equation}
	\tilde{f}(\vect{p}) \equiv  \sum_{\vect{z}\in (a \mathbb{Z})^d} e^{i\vect{p}\cdot\vect{z}} f_{\vect{z}} = \tilde{f}({\vect{0}}) \exp\left( - \frac{\sigma^2\vect{p}^2}{2} \right)  \, .	\label{kerfour}
\end{equation}
Thus, the smeared quark operator has maximal overlap with a quark at rest, $\vect{p}=\vect{0}$.
Non-zero velocities are suppressed in accordance with the above Gaussian momentum distribution.
Clearly, for hadrons carrying significant spatial momenta, such a smearing may be counterproductive.

Having identified the problem, it is easy to modify the smearing to perform well for moving hadrons.
We aim at a momentum distribution of quarks centred around a finite momentum $\vect{k}$, so we need to shift the smearing kernel in momentum space:
\begin{equation}
	\tilde{f}(\vect{p}) \mapsto  \tilde{f}(\vect{p} - \vect{k}) \label{kerfourshift}\,,
\end{equation}
as illustrated in Fig.~\ref{fig:transform}.
This translates to the replacement in position space
\begin{equation}
	f_{\vect{z}} \mapsto e^{i\vect{k}\cdot\vect{z}} f_{\vect{z}} \ .
\end{equation}
Our modified smearing operator $F_{(\vect{k})}$, where $F_{(\vect{0})}=F$,
can thus be formally expressed as
\begin{align}
	(F_{(\vect{k})} q)_{\vect{x}} &= \sum_{\vect{y}\in (a \mathbb{Z})^d} F_{(\vect{k})\vect{x}\vect{y}} q_{\vect{y}}\nonumber\\&= 
	\sum_{\vect{y}\in (a \mathbb{Z})^d} e^{-i\vect{k}\cdot(\vect{x}-\vect{y})}\, f_{\vect{x}-\vect{y}}\, \GC_{\vect{x}\vect{y}}\, q_{\vect{y}} \label{formnewsm}\ .
\end{align}
The only new ingredient is the additional
phase factor $\exp\left[-i\vect{k}\cdot(\vect{x}-\vect{y})\right]$.
Note that the quark momentum shift
$\vect{k}$ need \emph{not} be a lattice momentum, i.e.\ it is not restricted to discrete values $\vect{k}\in(2\pi/L)\mathbb{Z}^d$. The smearing kernel $f$
and the gauge dependent factor $\GC$ can be taken over from any existing smearing method. For an iterative smearing method,
the extra phase factor can easily be integrated into the elementary smearing step. Below we demonstrate this for the
example of Wuppertal smearing.
In principle, there could be additional effects
like a Lorentz contraction of the wave function. This
was studied, e.g., in Refs.~\cite{Roberts:2012tp,DellaMorte:2012xc},
and we will also address this possibility.

The new, modified smearing operator $F_{(\vect{k})}$ of Eq.~\eqref{formnewsm} inherits important properties
from the smearing operator $F_{(\vect{0})}$ it is based on.
If the unmodified smearing operator is self-adjoint, then so is the modified smearing operator $F_{(\vect{k})}$,
because from $f_{\vect{x}-\vect{y}} = f_{\vect{y}-\vect{x}}^*$ and $\GC_{\vect{x}\vect{y}} = \GC^\dagger_{\vect{y}\vect{x}}$ it follows that
$F_{(\vect{k})\vect{x}\vect{y}} = F^\dagger_{(\vect{k})\vect{y}\vect{x}}$.
The underlying smearing operator $F_{(\vect{0})}$ should transform according to an irreducible representation (irrep) of the cubic group $O_h$.
Usually, this will be the trivial $A_1$ representation
but the smearing operator can also be used to inject angular momentum
or to add non-trivial gluonic excitations~\cite{Lacock:1994qx}.
Obviously, $F_{(\vect{k})}$ with a momentum shift $\vect{k}\neq\vect{0}$
breaks the cubic symmetry. However, when used in conjunction 
with a momentum projection that selects hadrons with a momentum $\vect{p}$, then as long
as $\vect{k}\parallel \vect{p}$, the smearing operator remains within the
$A_1$ representation of the $O_h$ little group corresponding to
the momentum direction $\vect{p}$.

\section{Momentum smearing and hadronic two-point functions}
\label{sec:wuppertal}
We recapitulate Wuppertal smearing as a generic example of
an iterative smearing algorithm.
Then to clarify notations we discuss the standard construction
of hadronic two-point functions,
before generalizing the smearing by introducing a momentum shift as described
above. The discussion can easily be extended to incorporate generic hadrons and
$n$-point functions with $n\geq 2$. 
We conclude with a suggestion how to
further improve the method. This is relevant for fine lattices
where the iteration count becomes large. In the Appendix
we discuss how an additional Lorentz boost factor can be implemented.
\subsection{Wuppertal smearing}
\label{sec:wup}
The most prominent gauge covariant realization of a smearing function is
Wuppertal smearing~\cite{Gusken:1989ad,Gusken:1989qx}, where
$F=\Phi^n$, with $\Phi$ being defined as
\begin{equation}
(\Phi q)_{\vect{x}} = \frac{1}{1+2d\wupc} \left[q_{\vect{x}}+\wupc \sum_{j=\pm 1}^{\pm d} U_{\vect{x},j} q_{\vect{x}+\unitj}\right]\,. \label{fermsmear}
\end{equation}
Again $d=3$ denotes the dimension of space. The gauge
transporters
\begin{align}\label{eq:gausssmear}
U_{x,-j} = U^{\dagger}_{x-\unitj,j}\,,\quad U_{\vect{x},j}=U_{x,j}\quad\text{with}\,\, x_4\,\,\text{fixed}\,,
\end{align}
can also be spatially APE smeared
links~\cite{Falcioni:1984ei,Alexandrou:1990dq,Bali:2005fu}.
$\unitj$ denotes a vector of length of one lattice unit $a$, pointing
into the direction $j$. In Eq.~\eqref{fermsmear} we suppressed the index $x_4$
since the smearing is diagonal in time. $\wupc$ is a positive
constant and the arbitrary normalization factor $1/(1+2d\wupc)$
is introduced to avoid a numerical overflow for large iteration
counts $n$.

$\Phi$ (and by implication $\Phi^n$) is self-adjoint, a unit matrix
in spinor space and transforms according to the $A_1$ representation of $O_h$
or its little groups.\footnote{$O_h$ symmetry will be reduced to $C_{nv}$
for momenta along a lattice axis ($C_{4v}$), a spatial diagonal ($C_{3v}$)
or a planar diagonal ($C_{2v}$), see, e.g., Ref.~\cite{Hamermesh:100343}.}
The replacement $q\mapsto \Phi^n q$ will therefore not interfere with
the symmetry properties of any interpolator.

The smearing operator $\Phi$ is related to a
discretized covariant Laplacian $\Laplace$:
\begin{align}
\Phi\,q&=q+\frac{\wupc}{1+2d\wupc}a^2\Laplace q\,,\\
a^2\left(\Laplace q\right)_{\vect{x}}
&=-2d\, q_{\vect{x}}+\sum_{j=\pm 1}^{\pm d} U_{\vect{x},j} q_{\vect{x}+\unitj}\,.\label{eq:laplace}
\end{align}
Introducing a fictitious time $\tau=n\Delta\tau$ and defining
$q(\tau)=\Phi^nq(0)$, i.e.\ $q(\tau+\Delta\tau)=\Phi q(\tau)$,
results in the diffusion (or heat) equation
\begin{equation}
\frac{\partial q(\tau)}{\partial\tau}\approx
\frac{q(\tau+\Delta\tau)-q(\tau)}{\Delta\tau}=\alpha\Laplace q(\tau)\,,\label{eq:diffuse}
\end{equation}
where
\begin{equation}
\alpha=\frac{a^2}{\Delta\tau}\frac{\wupc}{1+2d\wupc}\,.
\label{eq:alphadef}
\end{equation}
This is solved by $q(\tau)\approx \exp\left(\alpha\tau\Laplace\right)q(0)$.
Starting from $\delta$-sources in position and colour space
$q_{\vect{x}}^a(0)=\delta_{\vect{x}\vect{0}}\delta^{ab}$,
assuming the free case $U_{\vect{x},j}=\mathds{1}_3$
and large distances $r=|\vect{x}|\gg a$, Eq.~\eqref{eq:diffuse}
is obviously solved by a Gaussian with
\begin{equation}
\sigma(\tau)=\sqrt{2\alpha\tau}=\sqrt{2na^2}\sqrt{\frac{\wupc}{1+2d\wupc}}\label{eq:width}
\end{equation}
being the square root of the variance,
i.e.\ the smearing corresponds to the smearing kernel Eq.~\eqref{eq:smgauss}.
Employing a parallel transporter within the covariant Laplacian
Eqs.~\eqref{eq:gausssmear} and \eqref{eq:laplace}
that is close to unity, like
$d$ dimensional APE smeared gauge links,
means that this
Gaussian shape is a good approximation, see, e.g., Ref.~\cite{Bali:2011rd}.

The diffusivity $\alpha$ obviously is maximal for large values of the
parameter $\wupc$
($\alpha\rightarrow \frac16 a^2/\Delta\tau$ for $\wupc\rightarrow\infty$).
For small values ($\alpha\approx \wupc\, a^2/\Delta\tau$),
the iteration count
$n$ to achieve a given smearing radius\footnote{Note that the root mean
squared radius in $d$ dimensions reads
$\sqrt{d}\sigma$. Also
note that the width of the gauge invariant combination
$q^{\dagger}_{\vect{x}}q_{\vect{x}}$ Eq.~\eqref{density}
is smaller by a factor $\sqrt{2}$.}
$\sqrt{3}\sigma$ is larger but the
resulting wave function will be smoother. Equation~\eqref{eq:width}
highlights that, to keep the smearing radius fixed in physical
units, the iteration
count needs to be increased in proportion to the square of the
inverse lattice spacing. Obviously,
this becomes computer time intensive towards the
continuum limit. We remark that at small lattice spacings
the diffusion equation~\eqref{eq:diffuse} should be
solved using a smarter method than iteratively applying the smearing operator
$\Phi$ defined in Eq.~\eqref{eq:gausssmear}. We will return to this
in Sec.~\ref{sec:improve}.

\subsection{Construction of two-point functions} 
As an example we discuss the construction of momentum projected
pion and nucleon two-point functions:
\begin{align}
C_{\pi}(\vect{p},t)
&=\langle \pi_{\vect{p}} (t) \pi^{\dagger}_{\vect{p}} (0)\rangle\,,\nonumber\\
C_{N}(\vect{p},t)&=
\langle N_{\vect{p}}^{\alpha} (t) \overline{N}_{\vect{p}}^{\beta}(0)\rangle \mathbb{P}^{\alpha\beta}\,,
 \label{interpol}
\end{align}
where $\mathbb{P}=\frac{1}{2}(\mathds{1}+\gamma_4)$ denotes a projector onto
positive parity.\footnote{Note that for $\vect{p}\neq\vect{0}$
this projector will not completely remove the negative parity contribution.
To improve on this, instead one can employ the oblique
projector~\cite{Lee:1998cx}
$\mathbb{P}_{\vect{p}}=\frac{1}{2}\{\mathds{1}+[m_{N^*}/E_{N^*}(\vect{p})]\gamma_4\}$,
where $m_{N^*}$ and $E_N^*$ denote mass and energy of the
nucleon's negative parity partner. Since this state is higher
in mass than the nucleon, for simplicity, here we abstain from attempting this.}
Without smearing the pion and nucleon interpolators
are local quark bilinears and trilinears:
\begin{align}
\pi_{\vect{p}}= a^3\sum_{\vect{x}}e^{i\vect{p}\cdot\vect{x}} \bar{u}_{\vect{x}}\gamma_5 d_{\vect{x}}\,,\\
N^{\alpha}_{\vect{p}} = a^3\sum_{\vect{x}} e^{i\vect{p}\cdot\vect{x}} \left[ u^t_{\vect{x}} C \gamma_5 d_{\vect{x}} \right]u^{\alpha}_{\vect{x}}\,,
\end{align}
where $C$ is the charge conjugation operator and $u_{\vect{x}}$ and
$d_{\vect{x}}$ annihilate up and down quarks, respectively, at the
spatial position $\vect{x}$.

For the pion the Wick contraction then gives
\begin{align}
C_{\pi}(\vect{p},t)
&= -L^3a^3\sum_{\vect{x}}
e^{i \vect{p}\cdot\vect{x}} \left\langle\bar{u}_x \gamma_5 d_{x} d^{\dagger}_0 \gamma_5
\bar{u}^{\dagger}_0\right\rangle\label{eq:contract} \\\nonumber
&=-L^3a^3\sum_{\vect{x}}\left\langle\tr\left[(d\bar{d})_{x0}\gamma_5
(u\bar{u})_{0x}\gamma_5\right]\right\rangle e^{i \vect{p}\cdot\vect{x}}\\
&=-L^3a^3\left\langle\sum_{\vect{x}}\tr\left(M^{-1\dagger}_{x0}M^{-1}_{x0}
\right)e^{i \vect{p}\cdot\vect{x}}\right\rangle\,,\nonumber
\end{align}
where $x=(\vect{x},t)$ and the trace is over spin and colour.
Momentum projection at the source is not
necessary, due to the translational symmetry of expectation values
and $L^3=N^3a^3$ denotes the three-volume that corresponds to this omitted sum.
$M$ is a Wilson-like lattice Dirac matrix
with the quark mass $m_u=m_d$ and we
have used\footnote{Note that the assignment
$\bar{d}=d^{\dagger}\gamma_4$ is convenient but
not consistent in Euclidean spacetime, where
the Dirac operator is $\gamma_5$- instead of
$(\gamma_4=\gamma_0)$-Hermitian.
This is the reason for the negative sign in Eq.~\eqref{eq:contract}.
In the simulation, we correct for this phase and obtain a
positive two-point function.} $d^\dagger\gamma_5\bar{u}^\dagger=d^\dagger
\gamma_5\gamma_4u=-\bar{d}\gamma_5u$, the Grassmann nature of the
quark field creation and annihilation operators and replaced
$\langle d\bar{d}\rangle_d=\langle u\bar{u}\rangle_u=M^{-1}$,
where the subscripts $d$, $u$ denote integrating out the respective
Grassmann field
on a given gauge background. In the last step we also
exploited $\gamma_5$-Hermiticity:
$\gamma_5M^{-1}\gamma_5={M^{-1}}^{\dagger}$.
For the nucleon one can easily work out an analogous expression that contains
$\epsilon^{abc}\epsilon^{a'b'c'}(M^{-1})_{x0}^{aa'}
(M^{-1})_{x0}^{bb'}(M^{-1})_{x0}^{cc'}$, where we have suppressed the
spinor indices and $a,b,c,a',b',c'\in\{1,2,3\}$ run over fundamental
colour.
Equation~\eqref{eq:contract} can now be evaluated, solving the linear systems
\begin{equation}
\sum_{x,\alpha, a}M^{\beta' b'\alpha a}_{x'x}S^{\alpha a \beta  b}_x=\delta_{x'0}\delta^{b\beta'\beta}\delta^{b'b}\label{eq:prop}
\end{equation}
for twelve $\delta$-sources
(for each source spin $\beta$ and colour $b$) to obtain the point-to-all
propagator $S$ with sink position, spin and colour indices
$x$, $\alpha$ and $a$, respectively. Then,
\begin{equation}
C_{\pi}(\vect{p},t)\propto\left\langle\sum_{\vect{x}}\tr\left(S^{\dagger}_xS_x\right)e^{i\vect{p}\cdot\vect{x}}
\right\rangle\,,\label{eq:picontract}
\end{equation}
where again $x_4=t$. The construction of the analogous nucleon
two-point function from three point-to-all propagators
is straightforward. Note that in that case no Hermitian adjoint
will appear.

Smearing is diagonal in Euclidean time (hence
we suppressed the $t$ dependence) and trivial in Dirac spin. 
So, obviously, the smearing operator $F=\Phi^n$ commutes with
any $\Gamma$ structure, however,
it does not commute with covariant derivatives that may appear
within the hadronic interpolator. 
Smearing can easily be implemented
at the source, replacing $S=M^{-1}\delta$
by $S^F=M^{-1}F\delta$, see Eq.~\eqref{eq:prop}.
Note that, due to the fact that
$F$ is a unit matrix in spinor space, 
the same smeared source can be employed for all four
spinor components.
To obtain a so-called
smeared-smeared two-point function, the argument of the sum
in Eq.~\eqref{eq:picontract} will usually be replaced by
$\tr[(FS^{F})^{\dagger}_{x}(FS^{F})_{x}]e^{i\vect{p}\cdot\vect{x}}$:
every source smearing requires new inversions while
sink smearing needs to be carried out on all time slices
of interest.\\

We remark that momentum sources have been used for quite some time, see, e.g.,
Refs.~\cite{Gockeler:1998ye,Brown:2012tm,Bali:2015msa}.
Injecting momentum into quark sources is
necessary (and has been done) in the context
of the one-end-trick of Refs.~\cite{Sommer:1994gg,Foster:1998wu},
where one usually employs colour diagonal (complex) $\mathbb{Z}_N$ or
$\textmd{U}(1)$ random sources. Generalizing this
to $\textmd{SU}(3)$ noise is in fact equivalent to using a non-gauge fixed
wall source~\cite{Billoire:1985yn}, which does not change expectation
values. Momentum was however also injected into gauge fixed wall sources
in Refs.~\cite{Detmold:2012wc,Basak:2012py}
(``color wave propagator''), favourably affecting not only
statistical errors but also ground state overlaps. While these latter
references share some of our motivation, the method presented
here is quite different. For instance, in our case
the hadron's total momentum still needs
to be injected at the source and $\vect{k}$ is not quantized.\\

Finally, we remark that the asymmetry of only carrying
out the position sum at the sink often is exploited
to reduce the statistical errors of heavy-light meson
correlation functions, by only smearing the heavy quark with
$F^2=FF$ at the source, instead of smearing each quark with $F$.
Clearly, it would be advantageous if for each momentum
smearing parameter of interest only the heavy quark propagator had
to be recomputed. Unfortunately, momentum smearing (as well as
traditional smearing) as
an operation will not commute with momentum projection, unless
$\vect{p}=\vect{0}$. Therefore, a different distribution of the smearing
between the quarks will result in a different
and not necessarily optimal ground state overlap.

\subsection{Momentum (Wuppertal) smearing}
\label{sec:newsmear}
As explained in Sec.~\ref{sec:basic},
if we inject a momentum $\vect{p}$ into the hadron, it may be
a good idea to distribute at least part of it among the quarks.
Their smearing functions should therefore ideally be centred around
a value $\vect{k}\neq\vect{0}$, where $\vect{k}$ is
some fraction of $\vect{p}$, to best resemble the
wave function of the physical state we wish to create.\\

The Fourier transform of a Gaussian,
centred about $\vect{k}$  in momentum space, reads
\begin{equation}
f^{\sigma}_{(\vect{k})\,\vect{x}-\vect{y}}\propto
\exp\left[-\frac{(\vect{x}-\vect{y})^2}{2\sigma^2}-i\vect{k}\cdot(\vect{x}-\vect{y})\right]\,.\label{eq:momsf}
\end{equation}
Similarly, momentum can be injected also into differently
shaped smearing functions, a possibility that we shall not explore
here.  $q(\tau)=F_{(\vect{k})}^{\sigma(\tau)}q(0)$
solves the heat equation with a constant drift term
\begin{equation}
\label{eq:drift}
\frac{\partial q(\tau)}{\partial\tau}=
\alpha\left(\vect{\nabla}+i\vect{k}\right)^2 q(\tau)\,.
\end{equation}

In analogy to the discussion of Sec.~\ref{sec:wup},
in the free case the above smearing function can iteratively be constructed
from ``momentum'' Gau\ss{} smearing steps,
$F_{(\vect{k})} =\Phi_{(\vect{k})}^n$, introducing a phase 
into Eq.~\eqref{fermsmear}:
\begin{equation}
(\Phi_{(\vect{k})} q)_{\vect{x}} = \frac{1}{1+2d\wupc} \left[q_{\vect{x}}+\wupc \sum_{j=\pm 1}^{\pm d} U_{\vect{x},j}e^{i\vect{k}\cdot\unitj} q_{\vect{x}+\unitj}\right]\,, \label{fermsmear2}
\end{equation}
where $\Phi_{(\vect{0})}=\Phi$.
One can easily show that the variance $\sigma^2(\tau)$ is still given
as in Eq.~\eqref{eq:width}. Moreover, as expected
$\Phi_{(\vect{k})}$ remains self-adjoint:

\begin{widetext}
\begin{align}
\Phi_{(\vect{k})\vect{x}\vect{y}}&=\frac{1}{1+2d\wupc}\left\{\delta_{\vect{x},\vect{y}}+
\wupc \sum_{j=1}^{3} \left[\delta_{\vect{x}+\unitj,\vect{y}}
U_{\vect{x},j} e^{i\vect{k}\cdot\unitj}
+\delta_{\vect{y}+\unitj,\vect{x}}U^{\dagger}_{\vect{y},j} e^{-i\vect{k}\cdot\unitj}
\right]\right\}\,,\nonumber\\
\Phi_{(\vect{k})\vect{y}\vect{x}}^{\dagger}
&=
\frac{1}{1+2d\wupc}\left\{\delta_{\vect{y},\vect{x}}+
\wupc \sum_{j= 1}^{3} \left[\delta_{\vect{y}+\unitj,\vect{x}}
U_{\vect{y},j}^{\dagger} e^{-i\vect{k}\cdot\unitj}
+\delta_{\vect{x}+\unitj,\vect{y}}U_{\vect{x},j} e^{i\vect{k}\cdot\unitj}
\right]\right\}
=\Phi_{(\vect{k})\vect{x}\vect{y}}\,.\label{eq:transpose}
\end{align}
\end{widetext}

We will use the same smearing for quarks and
antiquarks. For the $\pi^-$ meson this amounts to the replacements
$d\mapsto \Phi^n_{(\vect{k})}d$,
$u\mapsto \Phi^n_{(-\vect{k})}u$, $\bar{d}\mapsto \bar{d}\Phi^n_{(\vect{k})}$
and $\bar{u}\mapsto \bar{u}\Phi^n_{(-\vect{k})}$
within Eq.~\eqref{eq:contract}. Note that the $\Phi^n_{(-\vect{k})}$ smearing is needed
as in the contraction
with the momentum projector, $(\delta_{\vect{x}\vect{x}'}e^{i\vect{p}\cdot\vect{x}})
\Phi^n_{(\vect{k})\vect{x}\vect{y}}\Phi^n_{(\vect{k})\vect{x}'\vect{y}'}$,
transposing the ordering of the indices of the first smearing operator
gives $\Phi^n_{(\vect{k})\vect{x}\vect{y}}=\Phi^n_{(-\vect{k})\vect{y}\vect{x}}$
in the free case,
see Eq.~\eqref{eq:freec}.
Exploiting the Hermiticity of $\Phi^n_{\vect{k}}$, we then obtain
\begin{align}
C_{\pi}(t)
\propto\sum_{\vect{x}}&\left\langle\!\tr\!\left\{\left[\left(\Phi^n_{(-\vect{k})}M^{-1}\Phi^n_{(-\vect{k})}\right)_{x0}\right]^{\dagger}\right.\right.\nonumber\\
&\qquad\times\left.\left.\left(\Phi^n_{(\vect{k})}M^{-1}\Phi^n_{(\vect{k})}\right)_{x0}
\right\}\right\rangle e^{i \vect{p}\cdot\vect{x}}\,.
\end{align}
In contrast, for baryonic two-point functions, where
all quarks propagate in the forward direction, only $\Phi^n_{(+\vect{k})}$
will appear. 
The pion two-point function can now be constructed in analogy
to Eq.~\eqref{eq:picontract} from
\begin{equation}
C_{\pi}(\vect{p},t)\propto\left\langle\!\sum_{\vect{x}}\tr\!\left[
\left(\Phi^n_{(-\vect{k})}S_{(-\vect{k})}\right)^{\!\dagger}_{\!x}\!\!\left(
\Phi^n_{(\vect{k})}S_{(\vect{k})}\right)_{\!x}\right]\!e^{i\vect{p}\cdot\vect{x}}\!
\right\rangle\,,
\end{equation}
where the source-smeared
point-to-all propagator is defined as [see Eq.~\eqref{eq:prop}]
\begin{equation}
{S_{(\vect{k})}}^{\alpha a \beta  b}_x=
\sum_{x',\alpha',a',z,c}\!\!\!\!\!\left(M^{-1}\right)^{\alpha a\alpha' a'}_{x,x'}\!\!\delta^{\alpha'\beta}
{\Phi^n_{(\vect{k})}}_{x'z}^{a'c}
\delta_{z0}\delta^{cb}\,,
\end{equation}
or, in shorthand notation: $S_{(\vect{k})}=M^{-1}\Phi_{(\vect{k})}^n\delta$.
For mesons the above twelve linear systems need to be solved, both for
$\delta$-sources smeared with $\Phi_{(\vect{k})}^n$ and
with $\Phi_{(-\vect{k})}^n$, while for baryons smearing with
$\Phi_{(\vect{k})}^n$ is sufficient.

We remark that the momentum smearing Eq.~\eqref{fermsmear2} can very
easily be implemented by substituting the (APE smeared) transporters
$U_{\vect{x},j}\mapsto U_{\vect{x},j}e^{i\vect{k}\cdot\unitj}$
and then iterating the usual Wuppertal smearing
Eq.~\eqref{fermsmear} on these modified
$\textmd{U}(3)$ links.

\subsection{Free field investigation}
\label{freefield}
Having defined the smearing and how the contractions are to be
carried out, we are now in the position to address the question
what value of $\vect{k}$
should be chosen. Naively, one may expect $\vect{k}\approx \vect{p}/2$
and $\vect{k}\approx\vect{p}/3$ to be optimal for mesons and baryons,
respectively, that are composed of degenerate quarks and carry
a total momentum
$\vect{p}$. This is indeed what we will find here for
the non-interacting case.
The interacting case somewhat deviates from this as we will
see in Sec.~\ref{sec:results} below.

Setting $U_{\vect{x},j}=\mathds{1}$ within Eq.~\eqref{fermsmear2}
gives the free case smearing function Eq.~\eqref{eq:momsf}
for a large volume $L^3\gg\sigma^3$ and number of smearing iterations
so that $\sigma\gg a$.
Smearing the quark and antiquark annihilation operators
at the sink with $F_{(\vect{k})}$
and $F_{(-\vect{k})}$, respectively, we obtain
the momentum projected smeared pion interpolator,
\begin{widetext}\begin{align}
\pi_{\vect{p}}&\propto
\sum_{\vect{x}, \vect{y},\vect{y}^{\prime}}
\bar{u}_{\vect{y}'}f_{(-\vect{k})\vect{y}'-\vect{x}} e^{i\vect{p}\cdot\vect{x}}
f_{(\vect{k})\vect{x}-\vect{y}}\gamma_5d_{\vect{y}}\nonumber\\
&\propto\sum_{\vect{x}, \vect{y},\vect{y}^{\prime}}\exp\left[-\frac{(\vect{y}-\vect{x})^2+(\vect{y}'-\vect{x})^2}{2\sigma^2}\right]
 e^{i\vect{p}\cdot\vect{x}}e^{-2i\vect{k}\cdot\vect{x}} e^{i\vect{k}\cdot(\vect{y}+\vect{y}^{\prime})} \left[\bar{u}_{\vect{y}^{\prime}} \gamma_5 d_{\vect{y}}\right]\label{eq:freec}\\\nonumber
&\propto \exp\left[-\frac{\sigma^2}{2}(\vect{p}-2 \vect{k})^2\right]
\sum_{\vect{Z},\vect{\Delta}}
\exp\left[-\frac{\vect{\Delta}^2}{\sigma^2}\right]
e^{i\vect{p}\cdot \vect{Z}} \left[\bar{u}_{\vect{Z}+\vect{\Delta}} \gamma_5 d_{\vect{Z}-\vect{\Delta}}\right]\,,
\end{align}
\end{widetext}
where we have defined centre and relative coordinates
\begin{align}
\vect{Z} = \frac{\vect{y}+\vect{y}^{\prime}}{2}\,,\quad
\vect{\Delta} =\frac{\vect{y}^{\prime}-\vect{y}}{2}\,.
\end{align}
Note that the components of $\vect{Z}/a$ and $\vect{\Delta}/a$
can be integer or half-integer valued, subject to
the constraints $(Z_j+\Delta_j)/a\in\mathbb{Z}$.
The result of Eq.~\eqref{eq:freec} is indeed maximized for
$\vect{k}=\vect{p}/2$:
in the free case $\vect{k}=\vect{p}/2$ is
the optimal smearing parameter for mesons
with mass-degenerate valence quarks while we encounter an exponential
suppression in $\vect{p}^2$ for the conventional
$\vect{k}=\vect{0}$ smearing. The broader the wave function in coordinate
space, the more
important becomes the correct choice of $\vect{k}$ as one can
see from the first exponent in Eq.~\eqref{eq:freec} (as well as from
Eq.~\eqref{kerfour}).
Unsurprisingly, for baryonic interpolators $\vect{k}=\vect{p}/3$
would be the optimal choice.

We have demonstrated that in the free case $\vect{k}$ can be
interpreted as the momentum
carried by the smeared quark. If the quarks differ in mass
or, like for the example of the nucleon, the interpolator
is not symmetric with respect to the quark flavour,
injecting different $\vect{k}$-values into different quark fields
may be advisable. In the interacting case
the interpretation is not as straightforward:
neither is the interpolator directly related to any of the
usual definitions of a wave function nor will all
momentum be carried by the valence quarks.

\subsection{Non-iterative (momentum) smearing}
\label{sec:improve}
We propose replacing iterative smearing, where the iteration count
diverges with $a^{-2}$ towards the continuum limit,
by a more refined method.
In this context we show how
to introduce phase factors (and shape functions) into
the ``distillation'' (or Laplacian-Heaviside) method of
Ref.~\cite{Peardon:2009gh}.
Although we already present the basics of the method here, systematic
tests are yet to be completed.
Another natural extension,
which we will investigate numerically in Sec.~\ref{sec:results}, is to
Lorentz boost the smearing function.

We define eigenvectors $|v_{\ell}\rangle$ and eigenvalues $-\omega_{\ell}^2$
of a covariant Laplacian with (smeared) gauge transporters,
at a fixed Euclidean time:
\begin{equation}
\label{eq:eigen}
\left(\Laplace +\omega_\ell^2\right)|v_{\ell}\rangle=0\,.
\end{equation}
Since the Laplacian is self-adjoint, the bra-ket notation is convenient
in the present context.
For a lattice of $N^d$ points per time slice ($L=Na$), $3N^d$ 
linearly independent eigenvectors $|v_{\ell}\rangle$ with
components $v^{a}_{\ell\vect{x}}$ exist.
Such eigenvectors were for instance used in Ref.~\cite{Peardon:2009gh}.
The heat equation~\eqref{eq:diffuse} is solved by
\begin{align}
|q(\tau)\rangle&=\exp\left(\alpha\tau\Laplace\right)|q(0)\rangle\nonumber\\
&=\sum_{\ell}| v_{\ell}\rangle\exp\left(-\alpha\tau\omega_{\ell}^2\right)\langle v_{\ell}|q(0)\rangle
\,.\label{eq:noniter}
\end{align}
If we start from a $\delta$-function at $\tau=0$, this results in a Gaussian of
variance $\sigma^2=2\alpha\tau$.

To implement momentum smearing one
can easily replace $U_{\vect{x},j}\mapsto U_{\vect{x},j}e^{i\vect{k}\cdot\unitj}$
within the covariant Laplacian $\Laplace\mapsto\Laplace_{(\vect{k})}\sim
(\vect{\nabla}+i\vect{k})^2$ and then
recompute the eigenvectors and -values:
\begin{equation}
\label{eq:eigen2}
\left(\Laplace_{(\vect{k})} +\omega_{(\vect{k})\ell}^2\right)|v_{(\vect{k})\ell}\rangle=0\,.
\end{equation}
The natural generalization of Eq.~\eqref{eq:noniter} for moving
particles then reads:
\begin{equation}
|q_{(\vect{k})}(\tau)\rangle
=\sum_{\ell}| v_{(\vect{k})\ell}\rangle\exp\left(-\alpha\tau\omega_{(\vect{k})\ell}^2\right)\langle v_{(\vect{k})\ell}|q(0)\rangle
\,.
\label{eq:noniter2}
\end{equation}

The motivation for computing eigenvectors of the covariant
Laplacian is to truncate Eqs.~\eqref{eq:noniter} or \eqref{eq:noniter2}
at a value $\ell_{\max}$ where $\omega_{\ell_{\max}+1}^2>\omega_{\max}^2$.
As $\alpha\tau=\sigma^2/2$, achieving a suppression by a
factor $e^{-2}$ requires $\omega_{\max}\approx 2/\sigma$.
In general, the wider the smearing function in coordinate space
the less eigenvectors will be needed. The extreme opposite
limit $\alpha\tau=0\Rightarrow\sigma=0$ corresponds to the Laplacian-Heaviside
method proposed in Ref.~\cite{Peardon:2009gh}, where summing over all
eigenvectors will ultimately result in
a $\delta$-function in position space while truncating 
at some finite $\ell_{\max}$ value
gives a bell shape~\cite{Peardon:2009gh}. (In the free case
the modulus would be a sum of sines.) The same holds for
the sources suggested in Ref.~\cite{Brown:2012tm} that correspond to
sums of eigenvectors of the non-interacting Laplacian.

It is trivial to work out more details in the free case, where
one basically encounters a one dimensional problem.
For instance, setting $\omega= j (2\pi/L)\gtrsim 2/\sigma$,
where $j\in\mathbb{Z}\setminus\{0\}$
means
$|j|\gtrsim L/(\pi\sigma)$. Therefore, in $d$ dimensions
$\ell_{\max}\gtrsim 3\times[\pi^{d/2}/\Gamma(d/2+1)]\times[L/(\pi\sigma)]^d$, where the
factor $3$ is due to colour and the next factor is the volume
of a $d$ dimensional unit-sphere ($4\pi/3$ for $d=3$): the number of
required eigenvectors increases in proportion to the
spatial volume in physical units
but is independent of the lattice spacing $a$.
In contrast, in the case of iterative smearing, reducing the lattice spacing
at a fixed value of $\sigma$ increases the iteration count
in proportion to $a^{-2}$, independent of the volume.
For our smearing size $\sigma\approx0.45\units{fm}$ and a spatial volume
$L^3=(6\units{fm})^3$, which
would ensure $Lm_{\pi}>4$ even at the physical mass point, we obtain
$\ell_{\max}\approx 960$ while
for a $(3\,\units{fm})^3$ box
about 120 eigenvectors should suffice.
Therefore, satisfactory results in terms of
ground state overlaps appear to be within reach, employing moderately
large numbers of eigenvectors. This is at present under investigation.

Non-Gaussian shapes can easily
be modelled too, e.g., by multiplying in a ``free form'' weight
function, $\left(|v_{\ell}\rangle\langle v_{\ell}|\right)_{\vect{x}\vect{y}}
\mapsto \sum_{\vect{xy}}h_{\vect{x}\vect{y}}\exp\left[\alpha\tau(\vect{x}-\vect{y})^2\right]\left(|v_{\ell}\rangle\langle v_{\ell}|\right)_{\vect{x}\vect{y}}$,
in analogy to Ref.~\cite{vonHippel:2013yfa}.
However, the numerical complexity of this operation is
$\mathcal{O}(N^{2d})$ for each time slice and eigenvector. At the
source this can be reduced to $\mathcal{O}(N^{d})$
if the solution is only required
for a fixed $\delta$-source position
$|q(0)\rangle=|\delta_{\vect{y}_0}^a\rangle$
($\langle\vect{y},b|\delta_{\vect{y}_0}^a\rangle=(|\delta_{\vect{y}_0}^a\rangle)_{\vect{y}}^b=\delta_{\vect{y}\vect{y}_0}\delta^{ba}$).
Other, numerically less expensive options include
placing non-linear functions of $\omega^2_{\ell}$ in the exponents
of Eqs.~\eqref{eq:noniter} and \eqref{eq:noniter2}, employing
linear combinations of Gaussians or substituting the covariant
Laplacian by a different
operator within Eqs.~\eqref{eq:eigen} and \eqref{eq:eigen2}, e.g.,
introducing an anisotropy.

Once the eigenvectors Eq.~\eqref{eq:eigen} or
Eq.~\eqref{eq:eigen2} of the APE or HYP
smeared covariant Laplacian have been computed,
any smearing radius can be realized, replacing
$\alpha\tau$ by the targeted $\sigma^2/2$ value within the exponent of
Eq.~\eqref{eq:noniter} or Eq.~\eqref{eq:noniter2}.
Another potential advantage of the non-iterative smearing is
that, instead of solving for smeared sources
$|q_{(\vect{k})}(\tau)\rangle$ that have evolved from
a $\delta$-source $|q(0)\rangle=|\delta_{\vect{y}_0a}\rangle$,
one can also directly apply the inverse lattice Dirac operator to
the eigenvectors, thereby
constructing so-called perambulators~\cite{Peardon:2009gh}:
\begin{equation}
{\tau_{(\vect{k})}}^{\alpha\beta}_{m t\, \ell 0}=\left\langle v_{(\vect{k})m t}\left|\left(M^{-1}\right)^{\alpha\beta}_{t0}\right|v_{(\vect{k})\ell 0}\right\rangle\,.
\end{equation}
The inner product above is over spatial position and colour,
replacing these indices by the eigenvector labels
$\ell$ at the sink and $m$ at the source. The respective
Euclidean times are
explicitly shown as additional eigenvector (and perambulator) subscripts.
These perambulators can then be folded with the appropriate
eigenvalue Gaussians during the construction
of hadronic $n$-point functions.

We abstain from repeating here the steps outlined in
Ref.~\cite{Peardon:2009gh} how to construct hadronic two-point
functions from these perambulators as there is only one
difference: In the contractions over $\ell$ and $m$ weight factors
$\exp(-\sigma^2\omega_{(\vect{k})\ell 0}^2/2)$
and $\exp(-\sigma^2\omega_{(\vect{k})m t}^2/2)$ should be introduced,
where $\omega_{(\vect{k})m t}^2$ refers to the $m$th eigenvalue
of the negative covariant Laplacian defined on time slice $t$.

On one hand using perambulators will require a larger number of inversions
and computationally more expensive contractions
than starting from $\delta$-sources.
On the other hand, due to volume averaging, the statistical
errors of the perambulator method will be smaller and this method allows
for more flexibility in the subsequent construction of hadronic
$n$-point functions.
Whether the use of perambulators
or of traditional (smeared) point-to-all propagators is preferable
will therefore depend on the problem at hand and in particular on
the number of different hadrons we are interested in. We have
shown that the same quark smearing can be employed in both cases.

Finally, we remark that in certain situations recomputing the eigenvectors
for different values of $\vect{k}$
may be avoidable. In the interacting case, gauge covariant derivatives
$\nabla^2_j$ will depend on all spatial coordinates, including
$x_i$ with $i\neq j$, and will therefore not commute with each other.
However, our intuition was based on the free case and
also the spatially APE~\cite{Falcioni:1984ei}
or HYP~\cite{Hasenfratz:2001hp} smeared gauge covariant
transporters are close to unity.
Assuming translational invariance, the coordinates can be separated and
the components of an eigenvector read
$v^a_{\vect{x}}\propto\sin(\omega_1^ax_1)\cdots\sin(\omega_d^ax_d)$,
where $\omega_j^a=2\pi m_j^a/L$ and $m_j^a$ are integer valued.
The corresponding eigenvalue of the negative Laplacian
is given as $\omega^2=\vect{\omega}^2$ where
the frequencies for different colour components
are constrained: $\vect{\omega}^2:={\vect{\omega}^1}^2={\vect{\omega}^2}^2={\vect{\omega}^3}^2$. Defining
\begin{equation}
|\tilde{v}_{(\vect{k})\ell}\rangle=e^{-i\vect{k}\cdot\vect{x}}|v_{\ell}\rangle
\label{eq:phases}
\end{equation}
then gives
$\left[\left(\vect{\nabla}+i\vect{k}\right)^2+\omega_{\ell}^2\right]|\tilde{v}_{(\vect{k})\ell}\rangle=0$.

In the interacting case this will not exactly solve the heat equation
with drift~\eqref{eq:drift} but
the approximation $\omega_{(\vect{k})\ell}\approx\omega_{\ell}$,
$|v_{(\vect{k})\ell}\rangle\approx|\tilde{v}_{(\vect{k})\ell}\rangle$ should
be sufficient to construct a gauge covariant Gaussian shape
with the intended phase factors.
Note that the phases Eq.~\eqref{eq:phases} appear both
in the bra- and in the ket-vector of Eq.~\eqref{eq:noniter2},
such that only relative
phases between two spatial positions matter and the choice of the zero
point becomes irrelevant. The clear disadvantage of the explicit
multiplication by phase factors Eq.~\eqref{eq:phases} is that these
have to obey the lattice periodicity, i.e.\ $k_j\in (2\pi/L)\mathbb{Z}$.
However, such a restriction may be tolerable on large lattices.
Also introducing twisted fermionic boundary
conditions~\cite{Bedaque:2004kc,Sachrajda:2004mi} may
provide a way to increase the flexibility of the choice of $\vect{k}$.

\section{Lattice ensemble and dispersion relations}
\label{sec:simulate}
We study the new momentum smearing method on 200 effectively decorrelated
configurations of $N_f=2$ non-perturbatively improved
Wilson fermions with the Wilson gluon action, generated by QCDSF
and RQCD. This constitutes a subset of
ensemble IV of Ref.~\cite{Bali:2014gha}.  Note that in
hadron structure studies we typically employ several sources
on about 2000 configurations~\cite{Bali:2014gha}, i.e.\
the present statistics are very moderate. Nevertheless, we
will report meaningful signals for momenta as high as
$2.8\units{GeV}$. The parameter values
of this $32^3\times 64$ ensemble read
$\beta=5.29$, $\kappa=0.13632$, corresponding to the inverse
lattice spacing  $a^{-1}\approx 2.76\units{GeV}$ and the (finite volume)
pion mass $m_{\pi}\approx 295\units{MeV}$. This gives a spatial
extent $L\approx 2.29\units{fm}\approx 3.42/m_{\pi}$.
Note that realizing the physical pion mass at this lattice spacing,
while keeping $L\gtrsim 3.5/m_{\pi}$, requires $L>70a$. As we are aiming
at momenta exceeding the hadron masses in any case, we do not
expect qualitative changes of results towards smaller pion masses
and have chosen the present parameters as a sufficiently
realistic but still affordable compromise. 

The momentum on a finite cubic lattice with even numbers of
points $N=L/a$ in each spatial direction
can take the discrete values
\begin{equation}
\label{quantize}
\vect{p}=\frac{2\pi}{L}\vect{P}\,,\quad
P_i\in\left\{-\frac{L}{2a}+1,-\frac{L}{2a}+2,\cdots, \frac{L}{2a}\right\}\,.
\end{equation}
This means the smallest non-trivial $|\vect{P}|=|(1,0,0)|$ gives
a momentum $|\vect{p}|\approx 0.54\units{GeV}$ while
$\vect{P}=(3,3,3)$ corresponds to $|\vect{p}|\approx 2.82\units{GeV}$.

The pion and nucleon masses have already been determined with high
statistics in Ref.~\cite{Bali:2014nma} and read
\begin{equation}
\label{eq:masses}
am_{\pi}=0.10675(52)\,,\quad am_N=0.3855(45)\,.
\end{equation}
We will compare our pion and nucleon
ground state energies to expectations from continuum and lattice
dispersion relations, using these reference values.
The continuum dispersion relation reads
\begin{equation}
\label{eq:disperse}
aE=a\sqrt{m^2+\vect{p}^2}\,.
\end{equation}
In addition, we will compare the pion energies to
the lattice dispersion relation for
a free naively discretized scalar particle,
\begin{equation}
\label{eq:dispersepi}
\cosh(aE_{\pi})=\cosh(am_{\pi})+\frac{a^2}{2}\hat{\vect{p}}^2\,,\quad
\end{equation}
and the nucleon energies to the dependence expected 
for a free Wilson fermion with the Wilson parameter $r=1$, see,
e.g., Ref.~\cite{ElKhadra:1996mp},
\begin{equation}
\label{eq:disperseN}
\cosh(aE_N)=1+\frac{\left(e^{am_N}-1+a^2\hat{\vect{p}}^2/2\right)^2+a^2\overline{\vect{p}}^2}{2\left(e^{am_N}+a^2\hat{\vect{p}}^2/2\right)}\,.
\end{equation}
Above, we used the standard abbreviations
\begin{equation}
\hat{p}_{\mu}=\frac{2}{a}\sin\left(\frac{ap_{\mu}}{2}\right)\,,\quad
\overline{p}_{\mu}=\frac{1}{a}\sin(ap_{\mu})\,,
\end{equation}
and $\hat{\vect{p}}^2=\sum_j\hat{p}_j^2$,
$\overline{\vect{p}}^2=\sum_j\overline{p}_j^2$.
We remark that the mass parameters $m_0$ within the
naive propagators (e.g., for the scalar case:
$1/(m_0^2+\hat{p}_{\mu}\hat{p}_{\mu})$)
differ from the rest frame energies by lattice artefacts.
The above masses are defined to satisfy $m=E(\vect{0})$, as it should be.
Their conversions to
$m_0$ are given as
$m_{\pi}=2a^{-1}\sinh (am_0/2)$ and $m_N=a^{-1}\ln(1+am_0)$, respectively.

\begin{figure*}[ht]
\centerline{\includegraphics[width=0.95\textwidth]{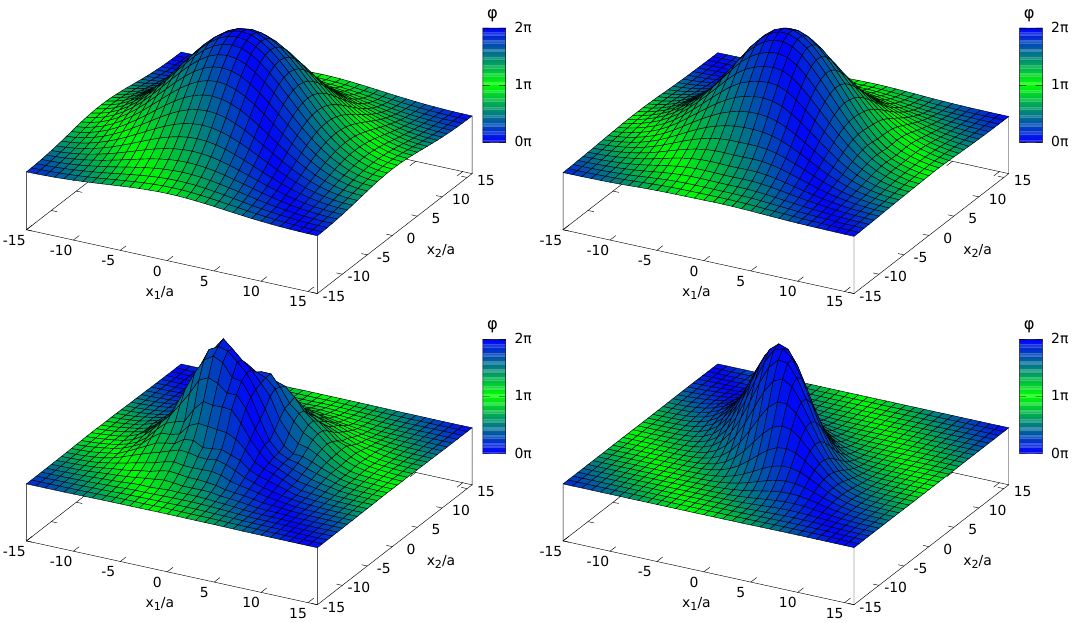}}
\caption{Cross sections of the smearing density profile $\rho(\vect{x})$
Eq.~\protect\eqref{density} in the $x_1$-$x_2$ plane.
The colour encodes the phase Eq.~\protect\eqref{phase}
and the momentum smearing $\vect{k}$ parameter has the direction $(1,1,0)$.
Top left: free field case. Top right: APE smeared gauge links.
Bottom left: original gauge links. Bottom right: APE smeared
links with an additional boost factor $\gamma=5.3$.}
\label{fig:smearing}
\end{figure*}

Obviously, the continuum dispersion relation should become
violated towards large momenta. In this case, we would not expect the
lattice dispersion relations, that apply to free pointlike particles,
to accurately describe the data either. However, the difference between
the continuum and the lattice formulae can serve as a naive estimate
of the expected size of lattice artefacts.

\section{Results}
\label{sec:results}
We first describe and check our implementation
of momentum Wuppertal smearing. 
Then, in Sec.~\ref{sec:momtest} we optimize the smearing parameters
and test the effectiveness of the method. In Sec.~\ref{sec:boo}
we investigate whether introducing an additional Lorentz contraction
is advantageous, before determining pion and nucleon dispersion relations
up to momenta of $1.94\units{GeV}$ and $2.82\units{GeV}$, respectively,
in Sec.~\ref{sec:dispersion}.
\subsection{Implementation of the smearing}
We iterate the momentum Wuppertal smearing on our
$d+1=3+1$ dimensional gauge ensemble described in Sec.~\ref{sec:simulate},
employing spatially APE smeared~\cite{Falcioni:1984ei} gauge
transporters within Eq.~\eqref{fermsmear2}. These are iteratively
constructed as follows.
\begin{equation}
\label{ape}
U_{\vect{x},i}^{(n+1)}= P_{\textmd{SU}(3)}\!\!\left(U_{\vect{x},i}^{(n)}+\apec\sum_{|j|\neq i}
U_{\vect{x},j}^{(n)}U^{(n)}_{\vect{x}+\unitj,i}U^{(n)\dagger}_{\vect{x}+\uniti,j}\right),
\end{equation}
where $i\in\{1,2,3\}, j\in\{\pm 1,\pm 2,\pm 3\}$:
the sum is over the four spatial ``staples'' surrounding
$U_{\vect{x},i}$ where, again, we suppressed the time index as the
smearing is local in time. $P_{\textmd{SU}(3)}$ is a gauge covariant
projector onto the $\textmd{SU}(3)$
group, defined by maximizing $\re\tr[A^{\dagger} P_{\textmd{SU}(3)}(A)]$.
We iterate over the three
diagonal $\textmd{SU}(2)$ subgroups to achieve this. Other
projection possibilities can, e.g., be found in
Refs.~\cite{Morningstar:2003gk,Bali:2005fu}.
We iterate Eq.~\eqref{ape}
15 times, using the weight factor $\apec=2.5$.

Momentum Wuppertal smearing
Eq.~\eqref{fermsmear2} is implemented, multiplying the APE smeared links
for a given $\vect{k}$ value by phases,
$U_{\vect{x},j}\mapsto U_{\vect{x},j}e^{i\vect{k}\cdot\unitj}$,
and then iterating the usual Wuppertal smearing
Eq.~\eqref{fermsmear} on these links. Within this article we set $\wupc=0.25$
to obtain smooth smearing functions at tolerable iteration counts.
Starting from three $\delta$-functions (one for each source colour $b$) at
the spatial position $\vect{0}$, we can define ``wave functions'' for
three different source colours,
\begin{equation}
\psi_{(\vect{k})\vect{x}}^{ab}=\sum_{\vect{x}',a'}\left(\Phi_{(\vect{k})}^n\right)_{\vect{x}\vect{x}'}^{aa'}\delta_{\vect{x}'\vect{0}}^{a'b}\,,
\end{equation}
and the associated gauge invariant density:
\begin{equation}
\label{density}
\rho(\vect{x})=\frac{\sum_{ab}|\psi_{(\vect{k})\vect{x}}^{ab}|^2}
{a^3\sum_{\vect{x},ab}|\psi_{(\vect{k})\vect{x}}^{ab}|^2}\,.
\end{equation}
$\rho(\vect{x})$ does not carry any information relating to the
$\textmd{U}(1)$ phases. We therefore define
\begin{equation}
\label{phase}
\varphi(\vect{x})=\arg\left(\frac{\psi_{(\vect{k})\vect{x}}^{11}}{
\psi_{(\vect{0})\vect{x}}^{11}}\right)\in[0,2\pi)\,,
\end{equation}
as the phase of momentum smearing,
relative to the standard Wuppertal smearing. Above, we have singled
out one particular colour component but any diagonal component
will give the same phase function $\varphi(\vect{x})$.
In the case of a free configuration, i.e.\ employing trivial
links $U_{\vect{x},i}=\mathds{1}$, Eqs.~\eqref{density}--\eqref{phase}
simplify since $\psi_{(\vect{k})\vect{x}}^{11}=\psi_{(\vect{k})\vect{x}}^{22}
=\psi_{(\vect{k})\vect{x}}^{33}$ and colour off-diagonal elements vanish.
Moreover, $\varphi(\vect{x})=\arg(\psi_{(\vect{k})\vect{x}}^{11})$ in this case.

In Fig.~\ref{fig:smearing} we display $\rho(\vect{x})$ in
lattice units for the two dimensional cross section $x_3=0$ and
different settings. The colour encodes the phase $\varphi(\vect{x})$.
We
employ $n=200$ momentum Wuppertal smearing iterations with
$\wupc=0.25$ and set $\vect{K}=(1,1,0)=[L/(2\pi)]\vect{k}$,
i.e.\ $\vect{k}$ lies within the depicted plane and its modulus
is given as $|\vect{k}|\approx 0.77\units{GeV}$. Up to discretization and
finite volume effects, with this smearing we expect the free
case variance $\sigma^2\approx (6.3a)^2$ from Eq.~\eqref{eq:width}, i.e.\
$\rho(|\vect{x}|=6.3a)\approx e^{-1}\rho(0)$. 
In the top left panel we show the free case result from our iterative
smearing, which is consistent with this expectation. In the top right
panel of Fig.~\ref{fig:smearing} we repeat this on one time slice
of one of our gauge configurations, after having APE smeared
the gauge links. The resulting shape is slightly narrower but
otherwise indistinguishable from the free case and almost invariant
with respect to continuous
rotations. However, rotations take
place in colour space: plotting individual components of
$q^{ab}_{\vect{x}}$ (not shown) gives a less smooth behaviour. In
particular, the off-diagonal components do not vanish anymore.
The smoothness of the gauge invariant density $\rho(\vect{x})$
means that the differences relative to the free case
can be removed almost completely by a suitable gauge transformation.

In the
bottom left panel we apply momentum Wuppertal smearing to a time slice of the
original, not APE smeared gauge links. In this case the resulting
density is less symmetric and less smooth.  As one
may expect from mean field
arguments~\cite{Parisi:1980pe}, the average
smearing radius is somewhat reduced in a way consistent with multiplying
$\wupc$ by the fourth root of the average plaquette. Also in
this case, the $\textmd{U}(1)$ phase information
is intact. A comparison with
the top right panel of Fig.~\ref{fig:smearing}
clearly demonstrates the advantage of additional gauge link
smearing.

Finally, 
in the lower right panel we apply boosted momentum Wuppertal
smearing Eqs.~\eqref{eq:boostsmear}--\eqref{boostnorm},
using the APE smeared gauge links. We set
$\gamma=5.3$, which corresponds to the ratio
of the pion energy for a momentum
$\vect{p}\approx 2\vect{k}$ over the pion mass $m_{\pi}=295\units{MeV}$.
The smearing parameter is converted
according to Eq.~\eqref{boosttrans}. Indeed, the perpendicular shape
in the central region is basically unaltered relative to the top panels
while in the direction parallel to the boost the density
is contracted by the $\gamma$ factor. Due to the large
numerical value of this factor there are slight deviations from
the theoretical expectation but
these discretization related effects can be removed by reducing the
smearing parameter $\wupc'$ and increasing the iteration count $n$,
keeping $\sigma$ constant, see
Eqs.~\eqref{eq:width} and \eqref{boosttrans}.

\subsection{Optimization and test of momentum smearing}
\label{sec:momtest}
We now compute smeared-point and smeared-smeared
pion and nucleon two-point functions at different lattice momenta
Eq.~\eqref{quantize}, where we denote momentum and smearing
vectors in physical units as $\vect{p}$ and $\vect{k}$, respectively,
and integer component lattice momenta as
$\vect{P}=[L/(2\pi)]\vect{p}$. Note that $\vect{K}=[L/(2\pi)]\vect{k}$
does not need to have integer valued components. We (mostly)
restrict ourselves to
\begin{equation}
\label{eq:zeta}
\vect{K}=\zeta\vect{P}\,,
\end{equation}
where the naive expectation would be
$\zeta=1/2$ for the pion and $\zeta=1/3$ for the nucleon,
see Sec.~\ref{freefield}.
From the resulting two-point functions, we define effective
energies
\begin{equation}
\label{effen}
E_{H,\mathrm{eff}}(\vect{p},t+a/2)=
\frac{1}{a}\ln\left[\frac{C_H(\vect{p},t)}{C_H(\vect{p},t+a)}\right]\,,
\end{equation}
where $H\in\{\pi, N\}$. For the non-perturbatively improved
action that we use, which contains a clover term that
couples adjacent time slices, the meaningful range of 
$t$ values is $t\geq 2a$, i.e.\ we plot effective energies, starting
at $2.5a\approx 0.18\units{fm}$.

\begin{figure}
\includegraphics[width=0.48\textwidth]{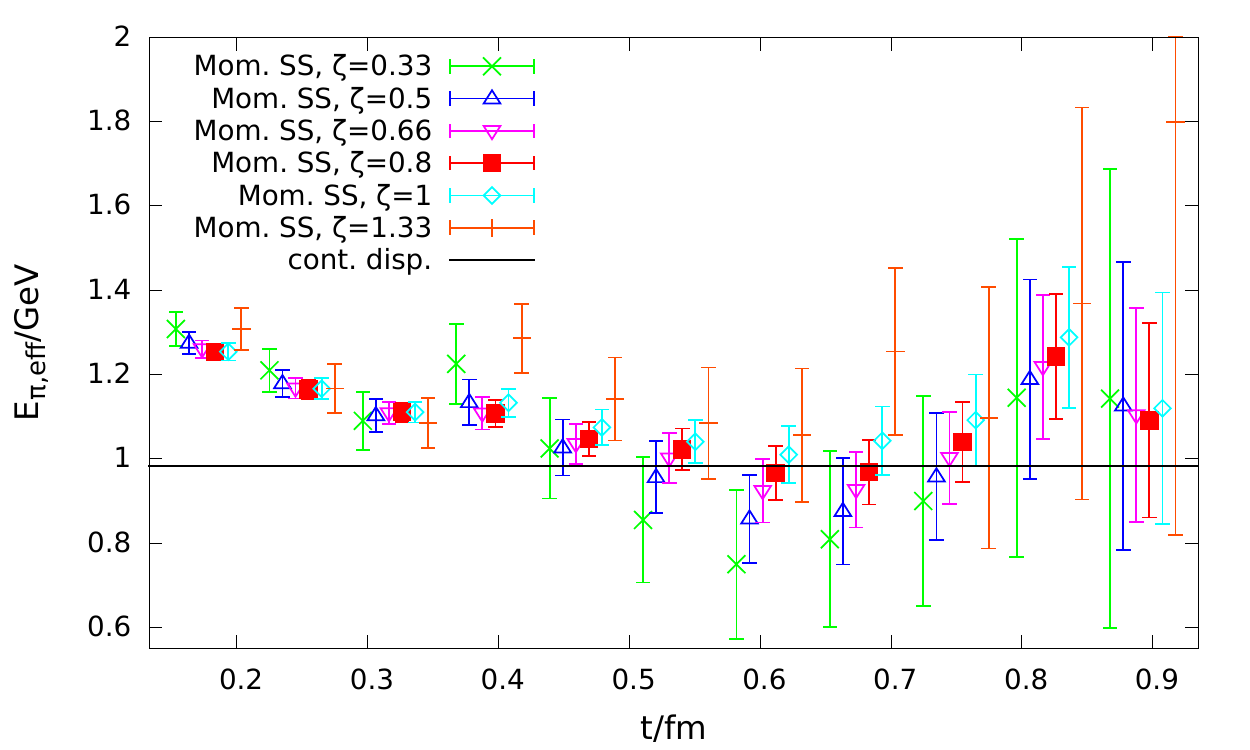}
\caption{Effective pion energies for $\vect{P}=(1,1,0)$, corresponding
to $|\vect{p}|\approx 0.94\units{GeV}$ and different ratios
$\zeta$, see Eq.~\protect\eqref{eq:zeta}.
The line is the expectation from
the continuum dispersion relation.
Symbols are shifted horizontally
to enhance the legibility.}
\label{fig:zeta}
\end{figure}

We realized different numbers of iteration counts $n$ both for momentum
and for conventional Wuppertal smearing. In both cases the best results
in terms of the ground state overlaps for pions and nucleons at different
momenta were obtained 
within the range $200\lesssim n\lesssim 400$. For the results we present here
we set $n=200$, $\varepsilon=0.25$, corresponding to
$\sigma\approx 6.3a\approx 0.45\units{fm}$, see Fig.~\ref{fig:smearing}.
We average the pion two-point function propagating in the forward
and backward time directions (folding). For the
nucleon we can only make use of the forward correlation function,
since we only employed the projector
$\mathbb{P}=\frac{1}{2}(\mathds{1}+\gamma_4)$,
see Eq.~\eqref{interpol}.

In Fig.~\ref{fig:zeta} we show effective energies from
smeared-smeared two-point functions for different values
of $\zeta$. In this case $\zeta=0.8$ (red squares)
appears to be the best choice, however the results are relatively
robust against increasing or decreasing this by 20\%.
In general,
at our pion mass $m_{\pi}\approx 295\units{MeV}$
and momenta up to $\sim 3\units{GeV}$, the optimal $\zeta$ values
came out to be $\zeta\approx 0.8>1/2$ for the pion and
$\zeta\approx 0.45>1/3$ for the nucleon. As discussed 
in Sec.~\ref{freefield}, we can only interpret
$\vect{k}$ as the momentum carried by a single
quark in the non-interacting case. Nevertheless, finding
values that are larger than the naive expectation,
rather than smaller, was somewhat unexpected.
Since the deviation from the free field case is not uniform
but bigger for the pion than for the nucleon, it would be
interesting to extend our study to non-Gaussian smearing functions.

\begin{figure}
  \includegraphics[width=0.48\textwidth]{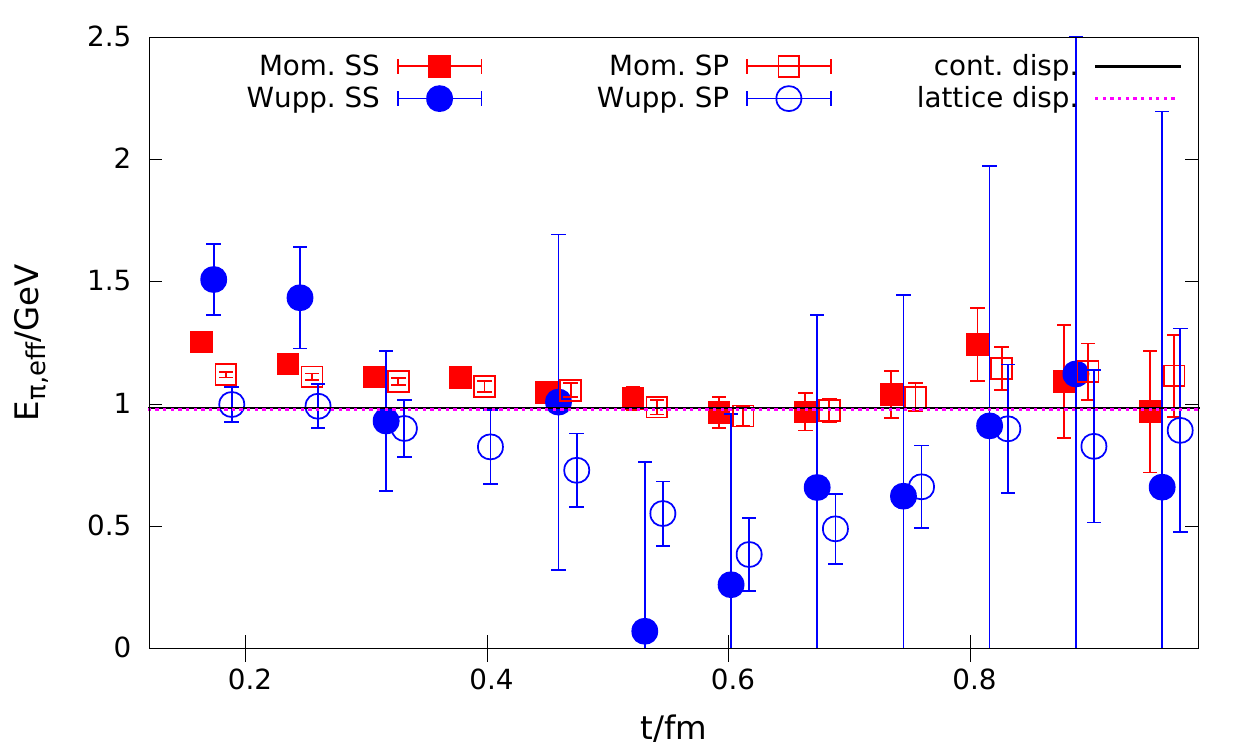}
\caption{Effective pion energies Eq.~\protect\eqref{effen}
for the lattice momentum
$\vect{P}=(1,1,1)$, corresponding to $|\vect{p}|\approx 0.94\units{GeV}$.
In the case of momentum smearing (squares), we set
$\vect{K}=\zeta\vect{P}$ with $\zeta=0.8$.
Solid symbols correspond to smeared-smeared, open symbols
to smeared-point two-point functions. Some data points
are shifted horizontally to enhance the legibility.
The expectations from
the continuum and lattice dispersion relations can be found in
Eqs.~\protect\eqref{eq:disperse} and \protect\eqref{eq:dispersepi}, respectively.
Symbols are shifted horizontally
for better legibility.}
\label{fig:mes1}
\end{figure}

\begin{figure}
  \includegraphics[width=0.48\textwidth]{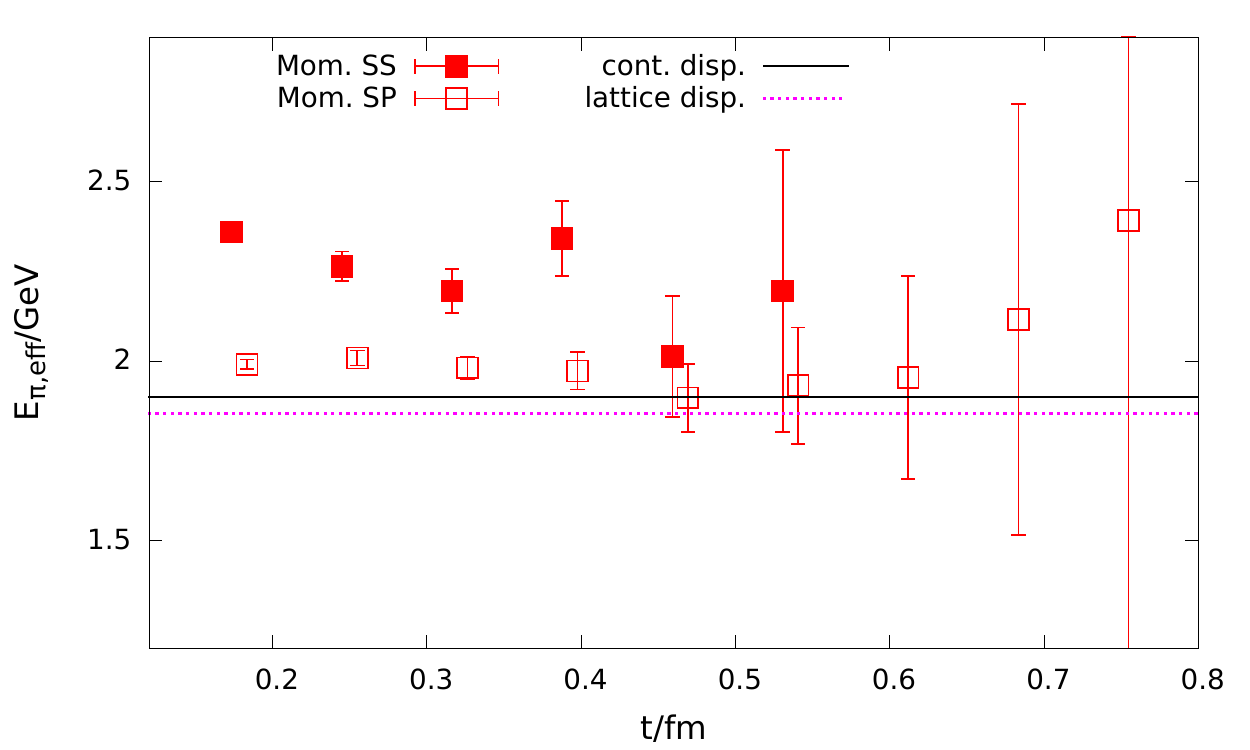}
\caption{The same as Fig.~\protect\ref{fig:mes1} for
$\vect{P}=(1,1,1)$, corresponding to $|\vect{p}|\approx 1.88\units{GeV}$.
The effective energies without momentum smearing cannot be determined,
due to prohibitively large errors and non-monotonous behaviour
of the central values of the respective two-point functions.}
\label{fig:mes2}
\end{figure}

\begin{figure}
  \includegraphics[width=0.48\textwidth]{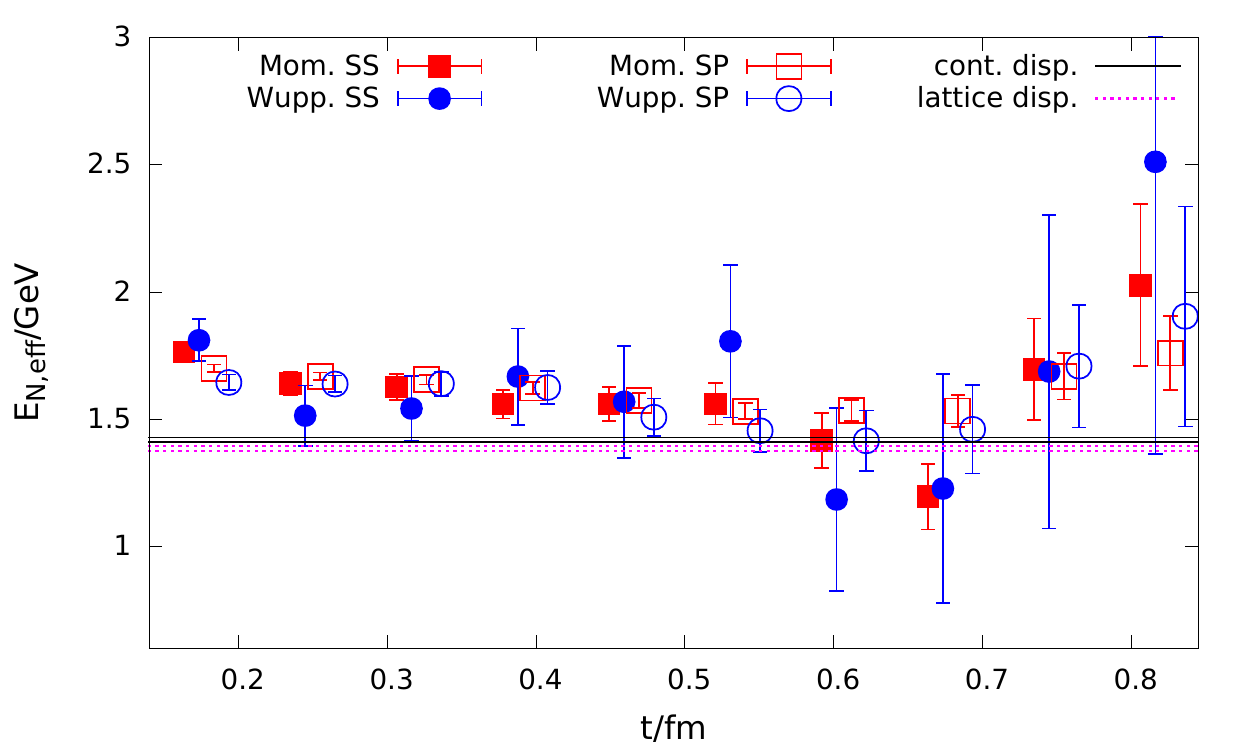}
\caption{The same as Fig.~\protect\ref{fig:mes1} for the nucleon
and $\zeta=0.5$. The horizontal lines correspond to the
expectations from the continuum and lattice dispersion relations
Eqs.~\protect\eqref{eq:disperse} and \protect\eqref{eq:disperseN}, respectively.}
\label{fig:nuc1}
\end{figure}

\begin{figure}
  \includegraphics[width=0.48\textwidth]{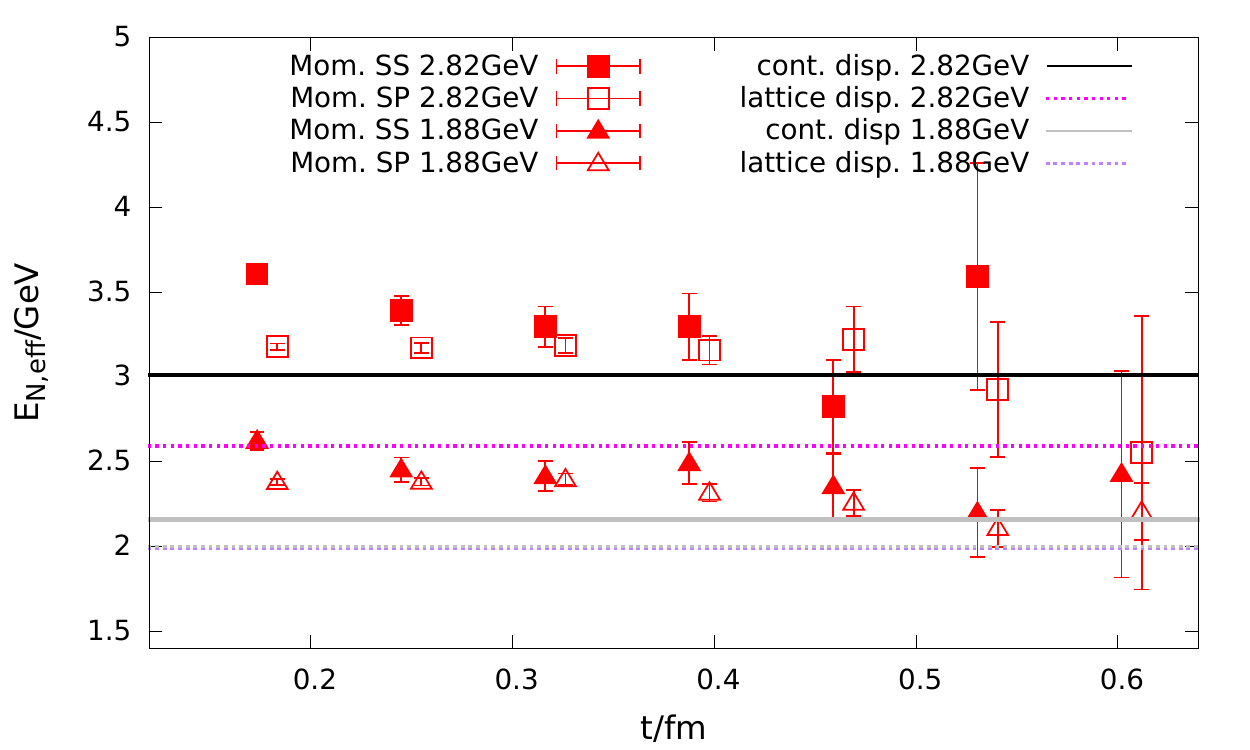}
\caption{The same as Fig.~\protect\ref{fig:nuc1} for 
$\vect{P}=(2,2,2)$ ($\zeta=0.4$) and $\vect{P}=(3,3,3)$ ($\zeta=0.44$),
corresponding to
$|\vect{p}|\approx 1.88\units{GeV}$ and $|\vect{p}|\approx 2.82\units{GeV}$, respectively.
Again, the conventional Wuppertal smearing data are too noisy
to allow for the extraction of effective energies.}
\label{fig:nuc2}
\end{figure}

In Figs.~\ref{fig:mes1} and \ref{fig:mes2} we compare effective pion energies
from smeared-point (SP) and smeared-smeared (SS) correlation functions
with the expectations from the continuum and lattice dispersion relations
Eqs.~\eqref{eq:disperse} and \eqref{eq:dispersepi},
using the pion mass Eq.~\eqref{eq:masses} (horizontal lines). Note that
the SS effective energies should be monotonous functions of $t$ while
this need not be the case for SP energies. It is well known
that statistical errors of SP correlators are smaller than
in the SS case and we also confirm this. For the momentum smeared
two-point functions the data are consistent with plateaus
for $t\gtrsim 0.5\units{fm}$ and we find agreement with the expectations.
For $|\vect{p}|\approx 0.94\units{GeV}$ (Fig.~\ref{fig:mes1}) also the effective
energies from
conventionally smeared two-point functions are consistent with the
expectation, however, the errors are too large to allow for
quantitatively meaningful statements. For $|\vect{p}|\approx 1.88\units{GeV}$
(Fig.~\ref{fig:mes2}), within our statistics of one source position on
200 gauge configurations, it turned out to be impossible to obtain
effective energies without momentum smearing at all.

The same comparison is shown for the nucleon in
Figs.~\ref{fig:nuc1} and \ref{fig:nuc2}, where in Fig.~\ref{fig:nuc2}
we also include a momentum as high as $|\vect{p}|\approx 2.82\units{GeV}$.
In this case we show the lattice dispersion relation
Eq.~\eqref{eq:disperseN} of a free Wilson fermion with the
nucleon mass given in Eq.~\eqref{eq:masses}.
The statistical errors of the momentum smeared data again
are much smaller than for the non-momentum smeared cases.
Like for the pion, it was impossible within our statistics
to extract effective energies for momenta larger than $1.5\units{GeV}$.
At $|\vect{p}|\approx 0.94\units{GeV}$ the data agree with the expectation
for $t\gtrsim 0.6\units{fm}$ while for the higher momenta, where
the statistical errors are larger, the data are consistent
with plateaus starting at $t\gtrsim 0.45\units{fm}$.
For the high momenta shown in Fig.~\ref{fig:nuc2}
the data seem to prefer the continuum dispersion relation over the
lattice dispersion relation Eq.~\eqref{eq:disperseN}.

\begin{figure}
\includegraphics[width=0.48\textwidth]{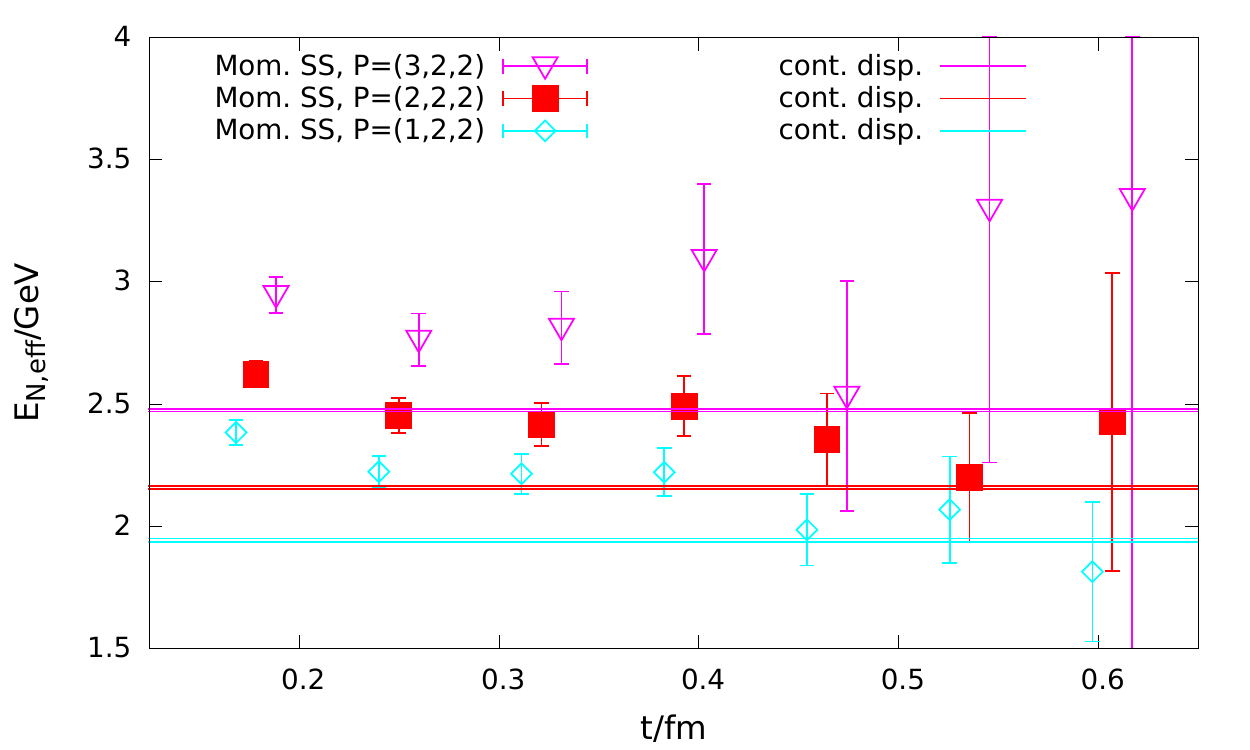}
\caption{The nucleon smeared-smeared effective energy with one
and the same momentum smearing $\vect{K}=(0.8,0.8,0.8)$,
optimized for $\vect{P}=(2,2,2)$, but for momenta pointing into
different directions.
Symbols are shifted horizontally
for better legibility.}
\label{fig:direction}
\end{figure}

It is clear from the results shown above that the gain of
using momentum smearing is tremendous. The only drawback, in addition
to the computational overhead from the smearing itself, is that
each value of the parameter $\vect{k}$ employed at the
source requires us to recompute the respective quark propagator.
A natural question therefore is whether it is possible
to efficiently realize several momenta $\vect{p}$ using one and
the same smearing vector $\vect{k}$.
We already know from Fig.~\ref{fig:zeta} that for
$\vect{p}\parallel \vect{k}$ the proportionality
constant $\zeta$ defined in Eq.~\eqref{eq:zeta} can be
varied by about 20\% without a significant deterioration
and by much more if one is willing to accept a compromise:
comparing Fig.~\ref{fig:zeta} with Fig.~\ref{fig:mes1}
reveals that even a much less than perfect momentum smearing
is a tremendous improvement over the conventional $\zeta=0$ case.
We may also ask whether it
is possible to (slightly) vary the direction of $\vect{p}$
relative to $\vect{k}$. This of course is potentially dangerous
since the interpolator used will not transform anymore
according to an irrep of the little group of $O_h$ (or its
double cover), associated to the momentum direction. However,
if for instance we are only interested in the ground state mass
of a spin-$0$ or spin-$1/2$ hadron this should not be a major problem.
Figure~\ref{fig:direction} demonstrates that to a certain extent varying
the momentum direction for a fixed $\vect{k}$
appears feasible too. In all the cases shown it is impossible to
extract any meaningful effective energies without momentum smearing.

\subsection{Test of boosted (momentum) smearing}
\label{sec:boo}
It has been suggested by two groups~\cite{Roberts:2012tp,DellaMorte:2012xc}
that introducing an anisotropy and thereby Lorentz boosting the smearing
function may improve the overlap of high momentum interpolators
with the respective hadronic ground states. We generalize
these ideas to off-axis momentum directions, also incorporating
phase factors, in the Appendix, see
Eqs.~\eqref{eq:boostsmear}--\eqref{boosttrans}.
Length contractions depend on the choice of coordinates
and in particular on time differences in the
moving frame relative to the rest frame. Since in Euclidean
spacetime all distances are spacelike and any real time
information is lost, it is not clear to us why spatial distances
should be subjected to a Lorentz boost.
Our numerical observations are negative.

\begin{figure}
\includegraphics[width=0.48\textwidth]{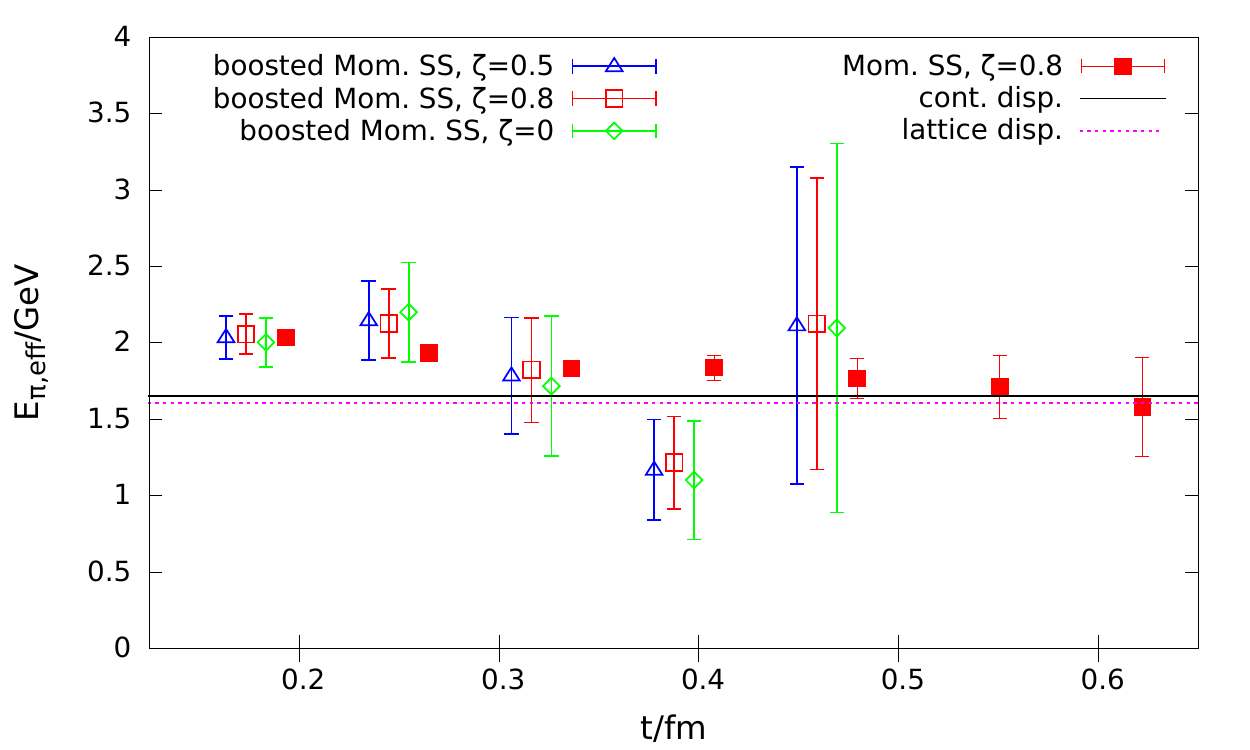}
\caption{Effective smeared-smeared
pion energies for $\vect{P}=(3,0,0)$, corresponding
to $|\vect{p}|\approx 1.63\units{GeV}$:
momentum smearing with the optimized parameter $\zeta=0.8$
and boosted momentum
smearing Eq.~\protect\eqref{eq:boostsmear} for $\gamma\approx 5.6$
with $\zeta=0.5, 0.8, 0$.
Symbols are shifted horizontally
for better legibility.}
\label{fig:boost}
\end{figure}

In Fig.~\ref{fig:boost}, which is representative for our experiences,
we show effective pion energies for $\vect{P}=(3,0,0)$,
corresponding to $|\vect{p}|\approx 1.63\units{GeV}$ and a boost factor
$\gamma=E_{\pi}(p)/m_{\pi} \approx 5.6$, close to the one
that corresponds to the bottom right panel of Fig.~\ref{fig:smearing}
($\gamma=5.3$). In this case it was again not possible to extract
effective energies using the conventional Wuppertal smearing.
All Lorentz contracted (boosted) interpolators give results much
inferior to the one obtained using the unboosted momentum smearing.
This also holds for different contraction factors $1/\gamma$ (not shown).
At the same time the boost enhances the signal relative to
unboosted Wuppertal smearing, without the momentum phase factor.
In the figure we show boosted smearing results for different momentum shifts:
$\vect{k}=\vect{0}$, the naive expectation $\vect{k}=0.5\vect{p}$ and
the case that is close to optimal without the boost applied,
$\vect{k}=0.8\vect{p}$. We find
the effective energies and their uncertainties
are quite insensitive to the $\vect{k}$ value. 
This is not surprising since the
support of the smearing function in the direction of the momentum is
quite small. At the same time this small support
may explain why the boost outperforms
conventional Wuppertal smearing as broad wave functions are disfavoured
at high momenta, unless the $\vect{k}$ vector is introduced,
see Eq.~\eqref{eq:freec}.

In summary, substantially contracting the smearing function
in the direction of the momentum ameliorates the phase mismatch discussed
in this article. Therefore, some improvement over the conventional
isotropic smearing case can be achieved.
However, only momentum smearing correctly accounts for this effect
and we see no indication that injecting a momentum alters the
optimal shape of the modulus of the smearing function Eq.~\eqref{density}.

\subsection{Comparison with dispersion relations}
\label{sec:dispersion}
Our main aim here was to demonstrate the effectiveness of
momentum smearing. For this purpose it was sufficient to
consider only one source position on 200 individual
gauge configurations. The present state-of-the-art, however, is
to realize multiple sources on ten times as many
configurations.
In the near future we will compute a multitude
of physically interesting observables with enhanced statistics.
The masses shown in Eq.~\eqref{eq:masses} were already obtained
with high statistics and in Figs.~\ref{fig:zeta}--\ref{fig:boost}
we have compared effective energies against the
continuum and lattice dispersion relations
Eqs.~\protect\eqref{eq:disperse}--\protect\eqref{eq:disperseN},
using these values. 

\begin{figure}[ht]
\includegraphics[width=0.48\textwidth]{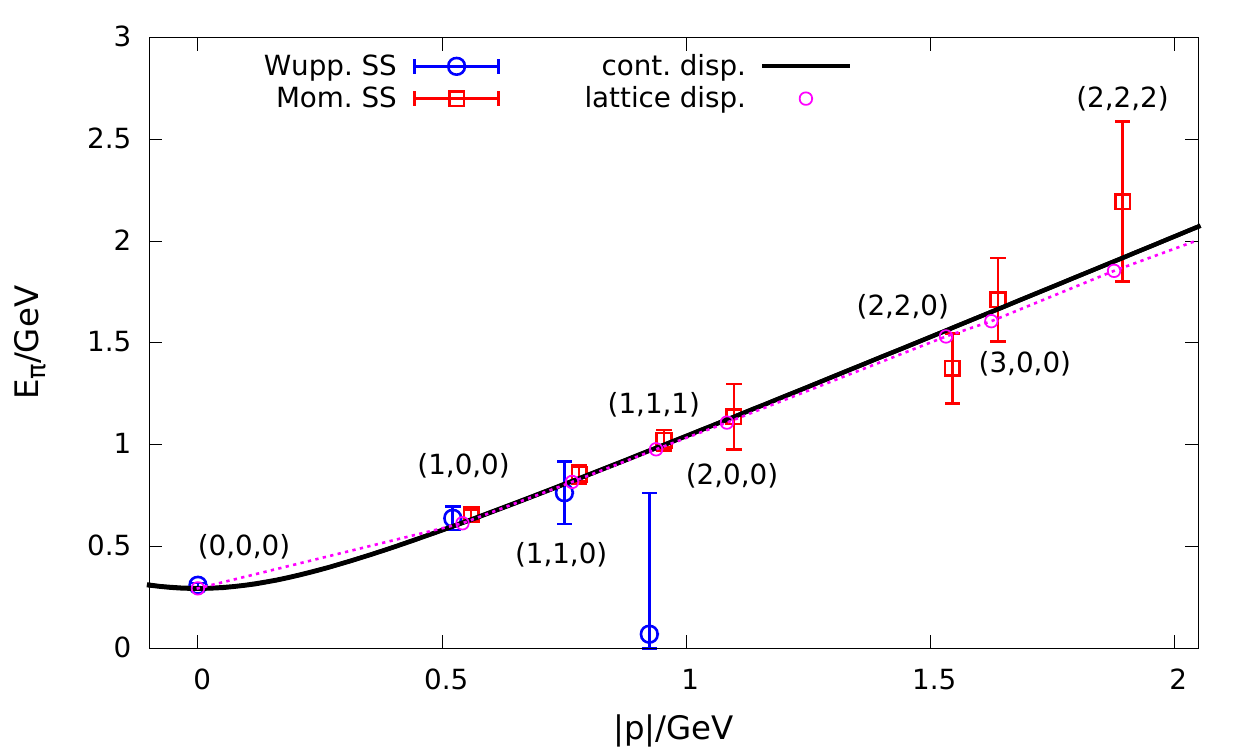}
\caption{Pion energies for different lattice momenta.
in comparison to the continuum (solid curve) and
lattice (points connected by dotted lines) dispersion
relations Eqs.~\protect\eqref{eq:disperse} and
\protect\eqref{eq:dispersepi}.} 
\label{fig:disppi}
\end{figure}

\begin{figure}[ht]
\includegraphics[width=0.48\textwidth]{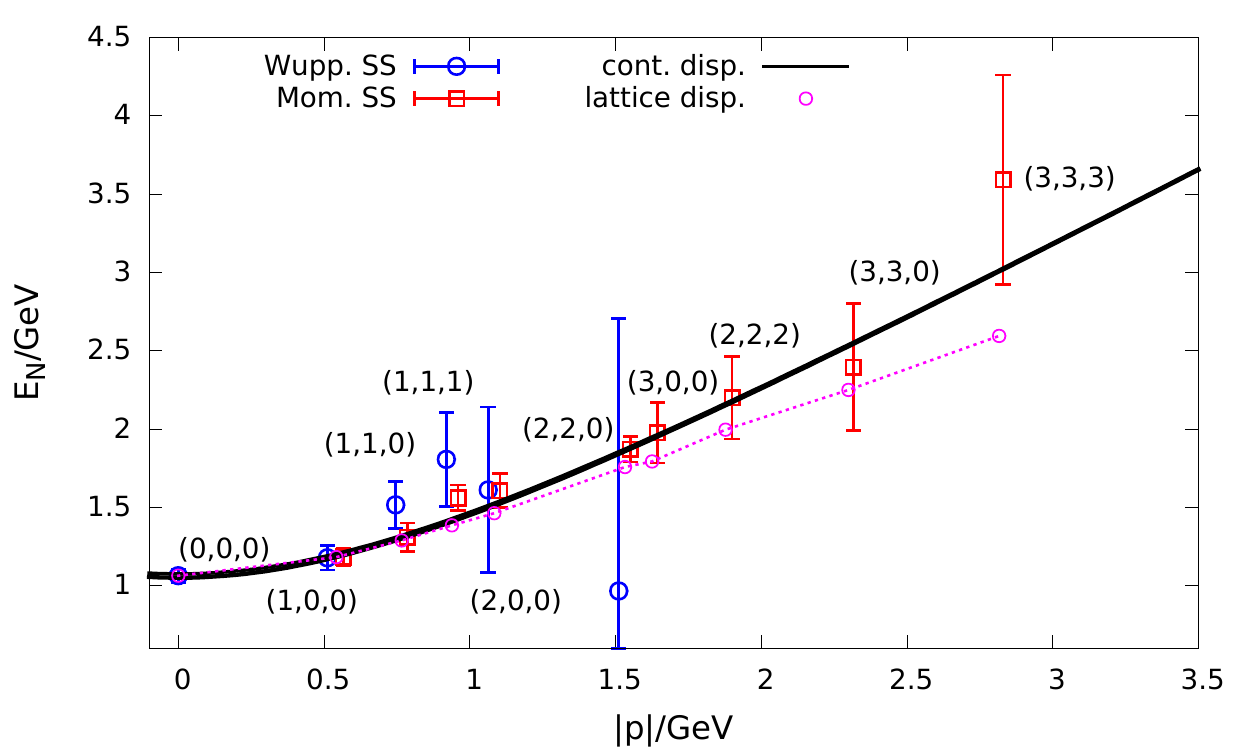}
\caption{Nucleon energies for different lattice momenta.
in comparison to the continuum (solid curve) and
lattice (points connected by dotted lines) dispersion
relations Eqs.~\protect\eqref{eq:disperse} and
\protect\eqref{eq:disperseN}.} 
\label{fig:dispN}
\end{figure}

In all cases the smeared-smeared effective energies from optimized
momentum smearing were in agreement with plateaus from
$t\geq t_{\min}=8.5a\approx 0.61\units{fm}$ onwards, where
$t=8.5a$ corresponds to the effective energy
obtained from the correlation function at
$8a$ and $9a$, see Eq.~\eqref{effen}. In many cases
$t_{\min}$ could be chosen smaller. For the moment being,
we conservatively approximate the energies by $E_H(\vect{p})\approx
E_{H,\mathrm{eff}}(\vect{p},t_{\min})$. The results as a function
of $\vect{p}$ are shown in Figs.~\ref{fig:disppi} and \ref{fig:dispN} and
compared to the dispersion relation expectations. We also display
results obtained with conventional smearing for small momenta where
this is possible. For the two-point functions studied here
the precision of the conventional results can be improved
at little computational overhead by averaging over (for the absolute momentum
values shown) six, eight or twelve equivalent directions.
We have not done this, to allow for a ``fair'' comparison of the
efficiency of the smearing methods. It is clear from the figures, however,
that the maximally possible error reduction, assuming
different momentum direction results to be
statistically uncorrelated, would not affect any of our conclusions.

We do not expect either parametrization shown in Figs.~\ref{fig:disppi} and
\ref{fig:dispN} to
perfectly describe the data as the lattice dispersion relations
are for point particles, assuming a particular form of the
effective Lagrangian. However, differences between the two functions
are indicative for the size of possible lattice effects.
While in the pion case differences between the parametrizations
are on the present level of statistics insignificant, the nucleon data
appear to be better described by the continuum dispersion relation.
In the near future we will further investigate this, increasing
our statistics and also employing a different smearing as described in
Sec.~\ref{sec:improve}.

\section{Conclusion}
In many lattice gauge theory applications hadrons carrying high
momenta are required. Due to the exponential increase of relative
errors of $n$-point functions with Euclidean time distances and diminishing
ground state sampling, high momenta previously were very difficult
or impossible to achieve. In Sec.~\ref{sec:basic}
we have introduced a new class of quark smearing methods
for the construction of hadronic interpolators that address and substantially
mitigate these problems. One particular realization of these
methods, that is trivial to implement and comes with very little computational
overhead, is momentum Wuppertal smearing, defined in Eq.~\eqref{fermsmear2}.
We tested this very successfully in Sec.~\ref{sec:momtest}, enabling us to
determine pion and nucleon energies for momenta as high as almost $2\units{GeV}$
and $3\units{GeV}$, respectively, with just 200 measurements, see
Figs.~\ref{fig:disppi} and \ref{fig:dispN}. These figures
also include a comparison
with the conventional method.

In Sec.~\ref{sec:boo} we investigated
the possibility of introducing an (additional)
Lorentz boost~\cite{Roberts:2012tp,DellaMorte:2012xc}. 
With and without a momentum phase factor included,
this gave some improvement over unboosted Wuppertal smearing,
possibly due to a dampening of the phase mismatch by
the more rapid fall-off of the interpolating wave function
in the direction of the momentum.
However, the results obtained were inferior to those of
momentum Wuppertal smearing, without any boost applied.
We conclude that the intuition of contracting the
wave function may be unjustified
since the hadron is moving in real time but not in imaginary
(Euclidean) time.

Iterative methods
(momentum smearing or not) suffer from high iteration
counts $n\propto a^{-2}$ as the continuum limit $a\rightarrow 0$
is approached, in addition to the naive volume factor due to an increasing
number of lattice sites. Therefore, in Sec.~\ref{sec:improve}
we introduce other non-iterative (momentum) smearing methods
that may be more suitable for small lattice spacings
and that also allow for the construction of non-Gaussian shapes.
These will be used by us in the near future.

Realizing momenta that are much larger than the hadron masses of
interest is of fundamental importance in several modern applications,
e.g., in direct determinations of (quasi) distribution
amplitudes~\cite{Aglietti:1998ur,Abada:2001if,Braun:2007wv},
of (quasi) (generalized) parton
distributions~\cite{Ji:2013dva,Ji:2015jwa,Lin:2014zya,Alexandrou:2015rja}
and moments of transverse momentum
distributions~\cite{Hagler:2009mb,Musch:2010ka,Musch:2011er,Engelhardt:2015xja}.
The new method allowed us to extract nucleon masses (employing very
moderate computational resources) up to momenta $\vect{p}^2\approx 7.9\units{GeV}^2$.
This clearly makes the above observables amenable to lattice simulations
in a realistic setting. For three-point functions, these methods
potentially even allow for virtualities
$Q^2=4\vect{p}^2\approx 30\units{GeV}^2$, switching a source momentum of
$-\vect{p}$ into a sink momentum of $+\vect{p}$.
The computation of these quantities with the new smearing
is, depending on the observable, either in progress or planned.\\

\acknowledgments
This research was funded by the Deutsche Forschungsgemeinschaft Grant
No.\ SFB/TRR 55. The authors benefited
from discussions with Vladimir Braun, Robert Edwards and Jian-Hui Zhang.
We also thank Nilmani Mathur, Sasa Prelovsek and Rainer Sommer
for comments relating to
an earlier version of the manuscript.
Gunnar~S.~Bali acknowledges the hospitality of the
Mainz Institute for Theoretical Physics (MITP) where a
part of this article was written.
We used the CHROMA~\cite{Edwards:2004sx} software package along with the
LibHadronAnalysis library
and the multigrid solver implementation
of Ref.~\cite{Heybrock:2015kpy} (see also Refs.~\cite{Richtmann:2016kcq,Heybrock:2014iga,Frommer:2013fsa}). Computations were performed on
Regensburg's QPACE~B Xeon~Phi system and on the
SFB/TRR~55 QPACE~2~\cite{Arts:2015jia} Xeon~Phi installation in Regensburg.
We thank Sara Collins,
Peter Georg, Benjamin Gl\"a\ss{}le and Daniel Richtmann
for their help and technical support.

\section*{APPENDIX: BOOSTED (MOMENTUM) SMEARING}
We describe how we introduce an anisotropy
into (momentum) Wuppertal smearing to introduce a
Lorentz contraction of the smearing
function~\cite{Roberts:2012tp,DellaMorte:2012xc}
along the momentum direction proportional to $1/\gamma$,
where in the continuum
\begin{equation}
\label{eq:gamma}\gamma=E(\vect{p})/m=\sqrt{1+\frac{\vect{p}^2}{m^2}}\geq 1\,.
\end{equation}
Note that $m$ above denotes the hadron mass and
the hadron momentum $\vect{p}$ differs from
the momentum smearing parameter $\vect{k}$. We remark that
as we are using equal Euclidean time
interpolators there is no compelling reason why such a boost
should be applied.

In order to ``boost'' the smearing function we need to replace
\begin{equation}
\vect{\nabla}+i\vect{k}\mapsto\left(\frac{\vect{\nabla}_{\parallel}}{\gamma}
+\vect{\nabla}_{\perp}\right)+i\left(\frac{\vect{k}_{\parallel}}{\gamma}+\vect{k}_{\perp}\right)
\end{equation}
within Eq.~\eqref{eq:drift}, where $\vect{\nabla}_{\parallel}
=\vect{e}(\vect{e}\cdot
\vect{\nabla})$, $\vect{\nabla}_{\perp}=\vect{\nabla}-
\vect{\nabla}_{\parallel}$ and
\begin{equation}
\vect{e}=\frac{\vect{p}}{|\vect{p}|}=\frac{\vect{k}}{|\vect{k}|}\,.
\end{equation}
For the Laplacian this means
\begin{equation}
\label{eq:lboost}
\Laplace\mapsto \frac{\vect{\nabla}_{\parallel}^2}{\gamma^2}
+\vect{\nabla}_{\perp}^2=\left(\frac{1}{\gamma^2}-1\right)\vect{\nabla}_{\parallel}^2+\Laplace\,,
\end{equation}
where
\begin{equation}
\vect{\nabla}_{\parallel}^2=\sum_{i,j}e_ie_j\frac{\partial}{\partial x_i}
\frac{\partial}{\partial x_j}\,.
\end{equation}
One can easily derive a corresponding
iterative smearing: Generalizing
Eqs.~\eqref{fermsmear}--\eqref{eq:alphadef} and \eqref{fermsmear2},
we obtain
\begin{widetext}
\begin{equation}
(\Phi^{\gamma}_{(\vect{k})} q)_{\vect{x}} = \frac{1}{N}\!\!\left\{\!q_{\vect{x}}
+\wupc'\!\left[\sum_{j=\pm 1}^{\pm d}\! N_{|j|}U_{\vect{x},j}e^{i\vect{k}\cdot\unitj} q_{\vect{x}+\unitj}+\left(\!1-\frac{1}{\gamma^2}\!\right)\!
\sum_{i\neq j=1}^d\!\!\frac{e_ie_j}{2}\left(U_{\vect{x},j}U_{\vect{x}+\unitj,-i}
+U_{\vect{x},-i}U_{\vect{x}-\uniti,j}\right)e^{i\vect{k}\cdot(\unitj-\uniti)}
q_{\vect{x}+\unitj-\uniti}\right]\!
\right\}\,,
\label{eq:boostsmear}
\end{equation}
\end{widetext}
where
\begin{equation}
N_j=
N_j(\gamma,\vect{e})=1+\left(\frac{1}{\gamma^2}-1\right)e_{j}\sum_{i=1}^de_i\,.
\end{equation}
The (arbitrary) normalization factor
\begin{equation}
\label{boostnorm}
N=N(\wupc',\gamma)=1+2\wupc'\left(d-1+\frac{1}{\gamma^2}\right)
\end{equation}
follows in the free case and is kept
to avoid numerical overflow for high iteration counts.
Like Eq.~\eqref{fermsmear2} this is most easily implemented,
replacing the (APE smeared) links $U_{\vect{x},j}$ by
$U_{\vect{x},j}e^{i\vect{k}\cdot\unitj}$ and then iterating
Eq.~\eqref{eq:boostsmear}, using these modified transporters,
instead of multiplying in the additional phase factors.

We remark that if the momentum is chosen parallel to
a lattice axis, the second sum within Eq.~\eqref{eq:boostsmear} 
vanishes. In this case the only differences with respect to the
momentum Wuppertal smearing defined in
Eq.~\eqref{fermsmear2} are that shifts in this direction carry a
suppression factor $1/\gamma^2$,
the normalization factor $N$ and $\wupc'\neq\wupc$.

The free field solution for $n$ iterations of
Eq.~\eqref{eq:boostsmear}
is a Gaussian with a phase factor
$e^{i\vect{k}\cdot\vect{x}}$ and variance,
\begin{equation}
\sigma^2=2na^2\frac{\wupc'}{N(\wupc',\gamma)}\,,
\end{equation}
in the directions perpendicular to the momentum, see Eq.~\eqref{eq:width}.
Parallel to the momentum the width is reduced by a factor $1/\gamma$,
as it should be. The perpendicular variance above can be kept fixed,
equating $\wupc'/N$ with $\wupc/(1+2d\wupc)$.
The resulting relation between the $\gamma=1$ smearing parameter $\wupc$
and the boosted parameter $\wupc'$ reads:
\begin{equation}
\label{boosttrans}
\wupc'=\frac{\wupc}{1+2\wupc\left(1-1/\gamma^2\right)}\leq\wupc\,.
\end{equation}

\bibliography{smear}
\end{document}